\documentclass[12pt]{article}

\newcommand{\drbar} {\overline{DR}}
\newcommand{\eq} [1] {Eq.~(\ref{#1})}
\newcommand{\fig} [1] {Fig.~\ref{#1}}
\newcommand{\msbar} {\overline{MS}}
\newcommand{\sect} [1] {Sect.~\ref{#1}}
\newcommand{\leer} [1] {}

\begin{document}

\title{
\begin{flushright} \small \rm
   ZU-TH 01/03 \\
  \small LC-TOOL-2003-042
\end{flushright}
       {\bf SPheno}, a program for calculating \\ supersymmetric spectra,  \\
       SUSY particle decays \\ and SUSY particle production at 
       $e^+ e^-$ colliders}
\author{W.~Porod\\[1cm] 
     Institut f\"ur Theoretische Physik, Universit\"at Z\"urich, \\
         CH-8057 Z\"urich, Switzerland}
\maketitle
\begin{abstract}
SPheno is a program that accurately calculates the supersymmetric
particle spectrum within a high scale theory, such as minimal
supergravity, gauge mediated supersymmetry breaking, 
anomaly mediated supersymmetry breaking, or string effective field theories.
An interface exists for an easy implementation of other high scale models.
The program solves the renormalization group equations numerically to 
two--loop order with user-specified boundary conditions.
The complete one--loop formulas for the masses are used which are supplemented
by two--loop contributions in case of the neutral Higgs bosons and the
$\mu$ parameter.
The obtained masses and mixing matrices are used to calculate
 decay widths and branching ratios
 of supersymmetric particles as well as of Higgs bosons,
$b \to s \gamma$, $\Delta \rho$ and $(g-2)_\mu$. Moreover, 
the production cross sections of all supersymmetric particle as well as
Higgs bosons at $e^+ e^-$ colliders can be calculated including initial
state radiation and longitudinal
polarization of the incoming electrons/positrons. 
The program is structured such that it can 
easily be extend to include non-minimal models and/or complex parameters.
\end{abstract}

\tableofcontents

\section{Introduction}

Supersymmetry (SUSY) \cite{Wess:tw,Nilles:1983ge,habkan} 
provides an attractive extension of the Standard Model 
(SM). It provides a qualitative understanding of various phenomena in
particle physics: It stabilizes
the gap between the Grand Unification scale / Planck scale and the
electroweak scale \cite{Witten:nf}. It allows the unification of the three
gauge couplings at a scale $M_U \simeq 2 \cdot 10^{16}$~GeV in a
straight forward way \cite{GUT}. The large top mass generates radiative 
electroweak symmetry breaking \cite{Ibanez:fr}. In addition it provides
the lightest supersymmetric particle as a cold dark matter candidate
 \cite{Ellis:1983ew}.
Therefore, the search for supersymmetric particles is one of the main topics in
the experimental program of present and future high energy colliders
\cite{Carena:1997mb,LHC,LC,Abdullin:1999zp}.

The Minimal Supersymmetric Standard Model (MSSM) consists of taking the
Standard Model and adding the corresponding supersymmetric partners
\cite{habkan}. In
addition a second Higgs doublet is needed to obtain an anomaly--free theory.
The second Higgs doublet is also needed to give mass to u-type quarks and
down-type quarks at the same time. 
The MSSM in its most general from contains
more than 100 unknown parameters \cite{Haber:1997if} which are clearly
to many for an exhaustive study.  This number drastically reduces if
one embeds the MSSM in a high scale theory, such as minimal supergravity 
theories \cite{sugra}, gauge mediated supersymmetry breaking\cite{gmsb},
or anomaly mediated supersymmetry breaking\cite{Giudice:1998xp}.
There is not yet a theoretical preferred scheme for supersymmetry breaking.
For this reason it is important to know whether the precision of
on-going and future experiments
is sufficient: (i) To distinguish between the various schemes. (ii)
To which extend it is possible to reconstruct the underlying theory.
It has been 
demonstrated that the expected experimental accuracies at future  $e^+ e^-$
colliders complemented with data from the LHC 
allow for a successful reconstruct of such an underlying
supersymmetric high scale theory \cite{Blair:2000gy}.  Connected with these
questions is the question if the theoretical accuracy matches the
experimental one. The present version of the  program  
\verb+SPheno+\footnote{SPheno stands for {\bf S}upersymmetric
{\bf Pheno}menology}
is thought as a further step in getting accurate
theoretical results to match finally the experimental precision.

In the view of ongoing and future experiments it 
is highly desirable to have various and independent tools at hand
performing the calculation of the supersymmetric spectrum, of decay widths,
of branching ratios and of production cross sections. This allows for a 
cross check of the tools and by comparing the implemented methods
and approximations one can also get a rough understanding of the
theoretical error. \verb+SPheno+ is a program performing an accurate
calculation of the supersymmetric spectrum, of the branching ratios of
supersymmetric particles and the Higgs bosons and of the production cross
sections of supersymmetric particles and the Higgs bosons 
in $e^+ e^-$ annihilation including longitudinal beam polarization. 
Moreover, the spectrum is used to calculate the branching of the rare
decay $b \to s \gamma$, the supersymmetric contributions to the
anomalous magnetic of the muon $a_\mu$ as well as supersymmetric
contributions to the $\rho$ parameter.  

 For the calculation of the spectrum the programs
 ISAJET \cite{Baer:1999sp}, SOFTSUSY \cite{Allanach:2001kg}
and SUSPECT \cite{suspect}  are widely used. A comparison of 
the results among these programs and with \verb+SPheno+
is given in \cite{Allanach:2002pz}. The calculation of the
branching ratios of supersymmetric particles as well as the production
cross sections in $e^+ e^-$ annihilation can  be done with \verb+SPheno+, 
ISAJET \cite{Baer:1999sp}, SPYTHIA \cite{Mrenna:1996hu},
SUSYGEN \cite{Ghodbane:1999va} and SDECAY \cite{Muhlleitner:2003vg}. 
A comparison of the results of these
programs will be given in a future paper.

\verb+SPheno+ has been written in Fortran90. The main focus has been
on accuracy and on stable numerical results and less on speed. However,
on a modern PC a typical running time is in the order of one second.
 The calculation is done using two-loop renormalization
group equations (RGEs) \cite{Martin:1993zk}, complete
one-loop correction to all SUSY and Higgs masses \cite{Pierce:1996zz}
supplemented by the 2-loop corrections  to the neutral Higgs bosons 
\cite{Degrassi:2001yf,Dedes:2002dy} 
and to the $\mu$ parameter\cite{Dedes:2002dy}.
The present version of \verb+SPheno+ does all calculations for real parameters
neglecting the flavour structure in the fermion as well as in the
sfermion sector. Decay widths and cross sections are calculated using
tree-level formulas. However, the couplings involved are running couplings
and thus important numerical effects of higher order corrections are
already taken into account.
The program
has been structured in such a way that the future inclusion of
complex phases and mixing between the generations has already been considered
in the design of the interfaces as well as in the definition of the various
variables. Moreover, extensions of the MSSM, e.g.~models with broken R-parity,
 can be implemented easily. 

The aim of this paper is to provide a manual of the program, version 2.0, 
to describe
the approximations used and to display the results of a run. 
In \sect{sect:MSSM} we will summarize the MSSM parameters and we give the
tree-level formulas for the supersymmetric particles. Moreover, a 
short summary of the implemented high scale models is given.
In \sect{sect:decays} we list the implemented decay modes of supersymmetric
particles and the Higgs bosons. We also discuss shortly 
the approximations used. 
In \sect{sect:prod} we present the implemented cross sections in
$e^+ e^-$ annihilation. In \sect{sec:constraints} we discuss give
details on the implemented low energy constraints.
In \sect{sec:calc} we discuss the implemented algorithm in some detail.
In \sect{sec:example} the main program is presented in detail providing
the necessary information so that this program can be easily adapted to
the user's requirement. In the appendices we discuss the possible
switches for influencing the program as well as a detailed discussion
of possible input files. This can be done using \verb+SPheno+ specific files
or by using the SUSY Les Houches Accord (SLHA) \cite{Skands:2003cj}.
 Moreover, we list the output of the program
for a typical example. The source code as well as precompiled a  version
of the program can be obtained from the author via email: 
\verb+porod@physik.unizh.ch+; or it can be downloaded from
\verb+http://www-theorie.physik.unizh.ch/~porod/SPheno.html+.

\section{MSSM parameters, particle spectrum, and Models}
\label{sect:MSSM}

In this section we fix our notation concerning the parameters and
present the tree-level formulas for the masses as well as the mixing matrices.
In the following we assume that the physical masses are ordered:
$m_i \le m_j$ if $i<j$ except for the sfermions as explained below.
We also give a short overview over various high scale models which
are implemented in the program.

\subsection{Ingredients for the Lagrangian}

The pure supersymmetric Lagrangian is specified by the K\"ahler potential
giving the gauge interactions
and by the Superpotential $W$ giving the Yukawa interactions:
\begin{eqnarray}
W = \epsilon_{ab} \left( Y^L_{ij} {\hat L}^a_i {\hat H}^b_1 {\hat E}^c_j 
  + Y^D_{ij} {\hat Q}^a_i {\hat H}^b_1 {\hat D}^c_j
  + Y^U_{ij} {\hat Q}^b_i {\hat H}^a_2 {\hat U}^c_j
  - \mu {\hat H}^a_1 {\hat H}^b_2\right)
\end{eqnarray} 
where ${\hat L}$, ${\hat E}$, ${\hat Q}$, ${\hat D}$, and ${\hat U}$
denote the matter superfields. The $SU(2)_L$ representation indices
are denoted by $a, b =1,2$ and the generation indices by $i,j=1,2,3$;
$\epsilon_{ab}$ is the totally antisymmetric tensor with $\epsilon_{12}=1$.
Note that the sign of $\mu$ is identical to the one in ISAJET 
\cite{Baer:1999sp} and SOFTSUSY \cite{Allanach:2001kg} but opposite 
to the convention in \cite{Pierce:1996zz}. Presently, real Yukawas
$Y^L$, $Y^D$, $Y^U$ only are included. They and the
gauge couplings $g_i$ are $\drbar$ quantities. $g_1$ is defined
in the Grand Unification normalization $g_1 = \sqrt{5/3} g'$ where
$g'$ is the Standard Model hypercharge gauge coupling.

The next ingredient is the soft SUSY breaking Lagrangian, which  is given by
mass terms for the gauginos
\begin{eqnarray}
L_{soft,1} &=& \frac{1}{2} \left( M_1 \tilde B \tilde B
              + M_2 \tilde W_a \tilde W^a
              + M_3 \tilde g_\alpha \tilde g^\alpha \right) + h.c. \, \, ,
\end{eqnarray}
mass terms for scalar matter fields and Higgs fields
\begin{eqnarray}
L_{soft,2} &=&  - M^2_{H_1} H^*_{1a} H^a_1 - M^2_{H_2} H^*_{2a} H^a_2
     - M^2_{\tilde L,ij} {\tilde L}^*_{ia} {\tilde L}^a_{j}
     - M^2_{\tilde E,ij} {\tilde E}^*_{i} {\tilde E}_{j}  \nonumber  \\
  && - M^2_{\tilde Q,ij} {\tilde Q}^*_{ia} {\tilde Q}^a_{j}
     - M^2_{\tilde U,ij} {\tilde U}^*_{i} {\tilde U}_{j}
     - M^2_{\tilde D,ij} {\tilde D}^*_{i} {\tilde D}_{j}
\end{eqnarray}
and trilinear couplings of scalar matter fields and Higgs fields
\begin{eqnarray}
L_{soft,3} &=& -
  \epsilon_{ab} \left( A^L_{ij} {\tilde L}^a_i  H^b_1 {\tilde E}^*_j 
  + A^D_{ij} {\tilde Q}^a_i  H^b_1 {\tilde D}^*_j
  + A^U_{ij} {\tilde Q}^b_i  H^a_2 {\tilde U}^*_j
  - B \mu  H^a_1  H^b_2\right) \nonumber \\ &&+ h.c.
\end{eqnarray}

\subsection{Masses and Mixing Matrices}

The masses of the various particles are induced by the soft SUSY
breaking parameters and the vacuum expectation values $v_i$ of the neutral
Higgs fields $v_i = <H^0_i>$. The ratio of the vacuum expectation values
is denoted by $\tan\beta = v_2 / v_1$. The sum of the vacuum expectation values
(vevs) squared is fixed by the gauge boson masses:
\begin{eqnarray}
m^2_W = \frac{1}{4} g^2 (v^2_1 + v^2_2) , \hspace{1cm}
m^2_Z = \frac{1}{4} (g^2 + {g'}^2) (v^2_1 + v^2_2)
\end{eqnarray}
Neglecting the mixing between different generations, the Standard Model 
fermion masses  are given by:
\begin{eqnarray}
m_{u_i} = \frac{1}{\sqrt{2}} Y^U_{ii} v_2 , \hspace{1cm}
m_{d_i} = \frac{1}{\sqrt{2}} Y^D_{ii} v_1 , \hspace{1cm}
m_{l_i} = \frac{1}{\sqrt{2}} Y^L_{ii} v_1
\end{eqnarray}
for u-quarks, d-quarks and leptons, respectively. 

The gluino mass is given by $m_{\tilde g} = |M_3|$. The charginos
are combination of the charged winos 
$\tilde w^\pm = (\tilde w^1 \mp i \tilde w^2) / \sqrt{2}$ and 
the charged higgsinos $\tilde h^-_1, \tilde h^+_2$. The Lagrangian
contains the chargino mass term 
$-(\tilde \psi^-)^T X \tilde \psi^+$ where
$\psi^- = (-i \tilde w^-, \tilde h^-_1)^T$, 
$\psi^+ = (-i \tilde w^+, \tilde h^+_2)^T$ and 
\begin{eqnarray}
X = \left( \begin{array}{cc} M &  \frac{g}{\sqrt{2}} v_2 \\
           \frac{g}{\sqrt{2}} v_1 & \mu \end{array} \right).
\end{eqnarray}
The matrix is diagonalized by two unitary matrices $U$ and $V$:
\begin{eqnarray}
M_{D,\tilde \chi^\pm} =  U^* X V^{-1} \, . 
\end{eqnarray}
The neutral gauginos $\tilde b, \tilde w^3$
as well as the
neutral higgsinos $\tilde h^0_1, \tilde h^0_2$ form the neutralinos. 
In the basis 
$\tilde \psi^0 = (\tilde b, \tilde w^3, \tilde h^0_1, \tilde h^0_2)^T$
one finds the mass term $- (\tilde \psi^0)^T Y \tilde \psi^0$ with
\begin{eqnarray}
Y = \left( \begin{array}{cccc}
         M_1 & 0   & - \frac{g'}{2} v_1 &   \frac{g'}{2} v_2 \\
          0  & M_2 & \frac{g}{2} v_1    & - \frac{g}{2} v_2 \\
      - \frac{g'}{2} v_1   & \frac{g}{2} v_1 & 0 & -\mu \\
        \frac{g'}{2} v_2 &   - \frac{g}{2} v_2    & -\mu & 0
 \end{array} \right)
\end{eqnarray}
This matrix is diagonalized by an unitary matrix $N$:
\begin{eqnarray}
M_{D,\tilde \chi^0} = N^* Y N^\dagger \,.
\end{eqnarray}

The CP-even electroweak eigenstates $(H^0_1, H^0_2)$ are rotated by the
angle $\alpha$ into the Higgs mass eigenstates $(h^0, H^0)$ as follows:
\begin{eqnarray}
\left( \begin{array}{c} h^0 \\ H^0 \end{array} \right) =
\left( \begin{array}{cc} -\sin\alpha & \cos\alpha \\
                          \cos\alpha & \sin\alpha \end{array} \right) 
\left( \begin{array}{c} H^0_1 \\ H^0_2 \end{array} \right) 
\end{eqnarray}
with $m_{h^0} < m_{H^0}$. The CP-odd and the charged Higgs masses are
given by
\begin{eqnarray}
 m^2_{A^0} = B \, \mu \, (\tan \beta + \cot \beta) \, ,
\hspace{1cm} m^2_{H^+} = m^2_{A^0} + m^2_W 
\end{eqnarray}
at tree level.

Neglecting generation mixing, the sneutrino masses are given by:
\begin{eqnarray}
M^2_{\tilde \nu_i} &=& M^2_{{\tilde L}_ii} + \frac{1}{2} m_Z^2 \cos 2 \beta
\end{eqnarray}
The other sfermion mass matrices are $2 \times 2$ matrices: 
\begin{eqnarray}
M^2_{\tilde l,i} &=& 
      \left( \begin{array}{cc}
        M^2_{\tilde L,ii} -\left(\frac{1}{2} - s^2_W \right) c_{2 \beta} m_Z^2
          + m_{l,i}^2 &
     \frac{1}{\sqrt{2}} \left( v_1 (A^L_{ii})^* - \mu Y^L_{ii} v_2 \right)   \\
     \frac{1}{\sqrt{2}} \left( v_1 A^L_{ii} - (\mu Y^L_{ii})^* v_2  \right)  &
        M^2_{\tilde E,ii} - s^2_W  c_{2 \beta}  m_Z^2  + m_{l,i}^2
          \end{array} \right) \\
M^2_{\tilde u} &=& 
      \left( \begin{array}{cc}
        M^2_{\tilde Q,ii} + \left(\frac{1}{2} - \frac{2}{3} s^2_W \right)
               c_{2 \beta}   m_Z^2 + m_{u,i}^2 &
    \frac{1}{\sqrt{2}} \left(  v_2 (A^U_{ii})^* - \mu Y^U_{ii} v_1\right)   \\
    \frac{1}{\sqrt{2}} \left(  v_2 A^U_{ii} - (\mu Y^U_{ii})^* v_1 \right)   &
        M^2_{\tilde U,ii} + \frac{2}{3} s^2_W c_{2 \beta}  m_Z^2 + m_{u,i}^2
          \end{array} \right) \\
M^2_{\tilde d} &=& 
      \left( \begin{array}{cc}
        M^2_{\tilde Q,ii} -\left(\frac{1}{2} - \frac{1}{3} s^2_W \right)
            c_{2 \beta}  m_Z^2  + m_{d,i}^2 &
   \frac{1}{\sqrt{2}} \left(   v_1 (A^D_{ii})^* - \mu Y^D_{ii} v_2  \right) \\
   \frac{1}{\sqrt{2}} \left(   v_1 A^D_{ii} - (\mu Y^D_{ii})^* v_2  \right)  &
        M^2_{\tilde D,ii} - \frac{1}{3} s^2_W  c_{2 \beta} m_Z^2 + m_{d,i}^2
          \end{array} \right) 
\end{eqnarray}
where $c_{2 \beta} = \cos \ 2 \beta$ and $s^2_W = \sin^2 \theta_W$.
These matrices are diagonalized by $2 \times 2$ matrices $R_{\tilde f,i}$ with
\begin{eqnarray}
m^2_{\tilde f} = R_{\tilde f} M^2_{\tilde f} R^\dagger_{\tilde f}
\end{eqnarray}
Sfermions are first ordered according to the generation and inside 
a generation according to their masses.
For example, in the slepton sector the ordering
is $\tilde e_1$, $\tilde e_2$, $\tilde \mu_1$, $\tilde \mu_2$, 
$\tilde \tau_1$, $\tilde \tau_2$ and similarly for squarks.

\subsection{High scale models}
\label{sect:models}

In this section we summarize the key ingredients of the high scale models
implemented in \verb+SPheno+. We also present the formulas for the boundary
conditions in the various models.
In all cases the modulus $|\mu|$ is determined
by requiring correct radiative symmetry breaking. At tree level the
corresponding formula reads as:
\begin{eqnarray}
|\mu|^2 = \frac{1}{2} \left[ \tan 2 \beta \left(
   M^2_{H_2} \tan\beta - M^2_{H_1} \cot\beta \right) - m^2_Z \right] \, .
\end{eqnarray}
Moreover, in all cases the high scale parameters are supplemented by the
sign of $\mu$ and $\tan\beta$.

\subsubsection{Minimal Supergravity}

The minimal supergravity (mSUGRA) scenario is characterized by a set
of universal parameters \cite{sugra,Nilles:1983ge} at the
GUT scale $M_{GUT}$: the gaugino mass parameter
 $M_{1/2}$, the
scalar mass parameter $M_0$, and the trilinear coupling $A_0$: 
\begin{eqnarray}
 M_i(M_{GUT}) &=& M_{1/2} \\
 M^2_{\tilde j}(M_{GUT}) &=& M^2_0 \\
 A_i(M_{GUT}) &=& A_0 Y_i(M_{GUT})
\end{eqnarray}

\subsubsection{Minimal Supergravity including right handed neutrinos}

In addition to the parameters of the mSUGRA model above the following 
parameters appear in this
case: $m_{\nu_R}$, a common mass for all right handed neutrinos, and
$m_{\nu_i}$ ($i=1,2,3$), the light neutrino masses. In this case
the MSSM RGEs are run up to the scale $m_{\nu_R}$ where the neutrino
Yukawa couplings are calculated using the formula 
$Y_{\nu,i} = \sqrt{m_{\nu_R} m_{\nu_i}}/v_2$. In the range between
$m_{\nu_R}$ and $M_{GUT}$ the effect of neutrino Yukawa couplings is
included in the RGEs of gauge and Yukawa couplings. At the GUT-scale
the right sneutrino mass parameters  as well as the trilinear coupling
$A_{\nu,i}$ are given by:
\begin{eqnarray}
 M^2_{\tilde R}(M_{GUT}) &=& M^2_0 \\
 A_{\nu,i}(M_{GUT}) &=& A_0 Y_{\nu,i}(M_{GUT})
\end{eqnarray}
 The corresponding RGEs are used in the running
from $M_{GUT}$ to $m_{\nu_R}$. At the scale $m_{\nu_R}$ the neutrino
Yukawa couplings $Y_{\nu,i}$, the trilinear couplings $A_{\nu,i}$
and the soft masses $M^2_{R,i}$ for the right sneutrinos are taken out
of the RGEs and below the  $m_{\nu_R}$ the usual set of MSSM RGEs are used.

\subsubsection{Gauge Mediated Supersymmetry Breaking}
\label{sect:gmsb}

Gauge mediated supersymmetry breaking \cite{gmsb,Giudice:1998bp} 
(GMSB) is characterized by the mass $M_M \sim$ $\langle S \rangle $ 
of the messenger
fields and the mass 
scale $\Lambda = {\langle F_S \rangle} / {\langle S \rangle}$ 
setting the size of the gaugino and scalar masses. 
The gaugino masses  
\begin{equation}
M_i(M_M) = (N_5+3 N_{10}) g\left(\Lambda/M_M\right)
\alpha_i(M_M) \Lambda
\label{bc1}
\end{equation}
are generated by loops of scalar and fermionic messenger component
fields; $N_i$ is the multiplicity of  messengers in the
$5+\overline{5}$ and $10+\overline{10}$ vector-like multiplets, and
\begin{eqnarray}
g(x) &=&{1+x\over x^2}\log(1+x)  +   (x\rightarrow-x)
\end{eqnarray}
is the messenger--scale threshold function \cite{Martin} which
approaches unity for $\Lambda \ll M_M$. Masses of the scalar fields in the
visible sector
are generated by 2-loop effects of gauge/gaugino and messenger fields:
\begin{equation}
M^2_{\tilde j} (M_M) = 2  (N_5 + 3 N_{10}) f\left(\Lambda/M_M\right)
\sum_{i=1}^3  k_i   C_j^i \alpha^2_i(M_M) \Lambda^2
\label{bc2}
\end{equation}
with $k_i=1,1,3/5$ for $SU(3)$, $SU(2)$, and $U(1)$, respectively;
the coefficients $C_j^i$ are the quadratic Casimir invariants,
being 4/3, 3/4, and $Y^2/4$ for the fundamental representations ${\tilde j}$ 
in the groups $i = SU(3), SU(2)$ and $U(1)$, with $Y=2(Q-I_3)$ denoting 
the usual hypercharge; also
the threshold function \cite{Martin} 
\begin{eqnarray}
f(x)&=& {1+x\over x^2}\biggl[\log(1+x) -2{\rm Li}_2
\left({x\over1+x}\right)  +\ {1\over2}{\rm Li}_2
\left({2x\over1+x}\right)\biggr] \nonumber \\ &&
  +   (x\rightarrow-x)
\end{eqnarray} 
approaches unity for $\Lambda \ll M_M$.
As evident from \eq{bc2} scalar particles with identical Standard--Model 
charges squared have equal
masses at the messenger scale $M_M$.
In the minimal version of GMSB, the $A$ parameters
are generated at 3-loop level and they are practically zero at $M_M$.
However, the program permits to set a value for $A_0$ different from zero
but universal for all sfermions.

\subsubsection{Anomaly Mediated  Supersymmetry Breaking}

In anomaly mediated supersymmetry breaking (AMSB) the SUSY breaking
is transmitted from the hidden sector to the visible sector via
the super--Weyl anomaly \cite{Giudice:1998xp}.
The soft SUSY breaking parameters are given by:
\begin{eqnarray}
 M_a &=& \frac{1}{g_a} \beta_a m_{3/2} \\
 A_i &=& \beta_{Y_i} m_{3/2} \\
 M^2_j &=& \frac{1}{2} \dot\gamma_j m^2_{3/2}
\label{eq:m0sqamsb}
\end{eqnarray}
where $\beta_a$ and $\beta_{Y_i}$ are the beta functions of gauge and Yukawa
couplings, respectively. The $\gamma_j$ are the anomalous dimensions of
the corresponding matter superfield and $m_{3/2}$ is the gravitino mass. 
Equation~(\ref{eq:m0sqamsb}) leads to
negative mass squared for the sleptons which is phenomenologically 
not acceptable. There are several possibilities to solve this problem
\cite{AMSBmodels} and the simplest one is to add a common scalar mass $M_0$ so
that eq.~(\ref{eq:m0sqamsb}) reads as 
\begin{eqnarray}
 M^2_j &=& M^2_0 + \frac{1}{2} \dot\gamma_j m^2_{3/2}
\label{eq:m0sqamsb1}
\end{eqnarray}
This extension has been implemented in the program.

\subsubsection{String Effective Field Theories}

Four--dimensional
strings naturally give rise to a minimal set of fields for inducing
supersymmetry breaking; they play the r\^ole of the fields in the hidden
sectors: the dilaton $S$ and the moduli $T_m$ chiral
superfields which are generically present in large classes of
4--dimensional heterotic string theories. The vacuum expectation
values of $S$ and $T_m$, generated by genuinely non--perturbative
effects, determine the soft supersymmetry breaking parameters 
\cite{cvetic,Binetruy:2001md}.

In the following we assume that all moduli fields get the same vacuum
expectation values and that they couple in the same way to matter fields.
Therefore, we omit the index $m$ and take only one
moduli field $T$.
 The properties of the supersymmetric theories
are quite different for dilaton and moduli dominated scenarios
as discussed in \cite{cvetic,Binetruy:2001md}.
The mass
scale of the supersymmetry parameters is set by  the gravitino mass
$m_{3/2}$.

In the program we implemented the complete 1-loop formulas given in
\cite{Binetruy:2001md}. Three classes of models are implemented 
in the program: two versions of $OII$ compactification defined
by the sets A and B of boundary conditions in  \cite{Binetruy:2001md}
as well as an $OI$ compactification scheme. For the implementation of
the   $OI$ compactification scheme we have used formulas Eqs.~(3.21) --
(3.23) of \cite{Binetruy:2001md}:
\begin{eqnarray}
M_i &=&  - g_i^2 m_{3/2}  \left\{  {\sqrt{3} \sin \theta} +
\left[ b_{i} + 
   {s \sqrt{3}\sin\theta} 
 g_{s}^{2}\left(C_i
-\sum_{j}C_{i}^{j}\right) \right. \right.
 \nonumber \\ && \hspace*{1cm} + \left. \left. 2 \, t
\cos\theta\,  G_{2}(t)
    \left( \delta_{\rm GS} + b_{i}
           - 2  \sum_{j}C_{i}^{j} (1+n_j)   \right)
 \right] / {16\pi^2}  \right\}  \\
M_{\tilde j}^2 &=& m^2_{3/2} \Bigg\{ \left(
  1 + n_j \cos^2 \theta \right) +  {2 \sqrt{3} s \sin\theta} 
 \left[ \sum_{i} \gamma_{j}^{i} g_{i}^{2} -
 \frac{1}{2s}
   \sum_{km} \gamma_{j}^{km} \right] \nonumber \\
    & & \hspace*{1cm} +
     \gamma_{j} + 2 t \cos \theta \, G_2(t)
  \sum_{km} \gamma^{km}_j (n_j + n_k + n_m+3) \Bigg\} \\
A_{jkm}&=& m_{3/2} \bigg[ - \sqrt{3} \sin \theta - 2 t
 \cos\theta  (n_j + n_k + n_m + 3) G_2(t) 
   + \gamma_j + \gamma_k + \gamma_m
 \bigg] \label{eq:a1}
\end{eqnarray}
$s = \langle S \rangle$ is the vacuum expectation values of the dilaton 
field.
$t = \langle T \rangle / m_{3/2}$ is the vacuum
expectation value of the moduli field(s), and 
$G_2(t) = 2\zeta(t) + 1/2t$ is the 
non-holomorphic Eisenstein function with
$\zeta$ denoting the Riemann zeta function.
$\delta_{GS}$ is the parameter 
of the Green-Schwarz counterterm.
$\gamma_j$ are the anomalous 
dimensions of the matter
fields, the $\gamma^i_j$ and $\gamma^{km}_j$ are their gauge and Yukawa parts,
respectively.
$C_i$, $C^j_i$ are the quadratic Casimir operators for the gauge group 
$G_i$, respectively, in the adjoint representation and in the matter 
representation. The indices $i,j,k$ denote $H_1$, $H_2$, $\tilde E$, 
$\tilde L$, $\tilde D$, $\tilde U$ and $\tilde Q$. The A-parameters are
finally given by:
\begin{eqnarray}
 A_{e,n}(GUT) &=& Y_{e,nn}(GUT) A_{\tilde E_n \tilde L_n H_1} \\ 
 A_{d,n}(GUT) &=& Y_{d,nn}(GUT) A_{\tilde D_n \tilde Q_n H_1} \\ 
 A_{u,n}(GUT) &=& Y_{u,nn}(GUT) A_{\tilde U_n \tilde Q_n H_2}
\end{eqnarray}
where $n$ denotes the generation.

 In case of the $OII$ compactification
scheme the gaugino masses are given by  Eqs.~(3.11) of \cite{Binetruy:2001md}:
\begin{eqnarray}
M_i&=&- g_i^2 m_{3/2} \Bigg\{ \frac{\sqrt{3} \sin\theta}
                                 {2 k_{s\overline{s}}^{1/2}} +
 \frac{1}{16 \pi^2} \bigg[  2 t \cos\theta G_{2}
                           \left( \delta_{\rm GS} + b_i \right) + b_i
 \nonumber \\
 && \hspace*{4.7cm} + 
\frac{\sqrt{3} g_s^2 \sin\theta}{2 k_{s\overline{s}}^{1/2}} ( C_i
-\sum_{j}C_{i}^{j} ) \bigg] \Bigg\}\, .
\label{MaO2}
\end{eqnarray}
For the sfermion parameters we have implemented two sets of boundary
conditions: set $(A)$ is specified by formulas Eqs.~(3.15) and (3.19)
of \cite{Binetruy:2001md}:
\begin{eqnarray}
M_i^{2} &=&m^{2}_{3/2} \Bigg\{ \sin^{2}\theta + \gamma_{i} +
\frac{\sqrt{3}\sin\theta }{k_{s\overline{s}}^{1/2}}
    \bigg[ \sum_{a} \gamma_{i}^{a} g_{a}^{2} +
   \frac{1}{2} 
   \sum_{jk} \gamma_{i}^{jk} (k_{s} + k_{\overline{s}}) \bigg] \Bigg\},
\label{massO2A} \\
A_{ijk} &=& m_{3/2} \Bigg\{ \gamma_{i} + \gamma_{j} + \gamma_{k}
   - \frac{\sqrt{3} k_s \sin\theta}{ k_{s\overline{s}}^{1/2}} \Bigg\}
\label{eq:a2}
\end{eqnarray}
Set $(B)$ is specified by  formulas Eqs.~(3.16) and (3.20)
 of \cite{Binetruy:2001md}:
\begin{eqnarray}
M_i^2&=&m^2_{3/2} \Bigg\{
\frac{\sqrt{3}\sin\theta}{k_{s\overline{s}}^{1/2}}
   \bigg[ 1 + 2 t \cos\theta G_{2} \bigg] \bigg[ \sum_{a} g_{a}^{2}
\gamma_{i}^{a} + \frac{1}{2} \sum_{jk} \gamma_{i}^{jk} \left(k_{s} +
k_{\overline{s}}\right) \bigg]  \nonumber \\
 & & \hspace{1cm}
   + \sin^{2} \theta \bigg[ 1 + \gamma_{i} + \ln\big[2 t 
     |\eta(t)|^4\big] \big( \sum_{a} \gamma_{i}^{a} +2 \sum_{jk}
   \gamma_{i}^{jk} \big) - \sum_{a} \gamma_{i}^{a} \ln(g_{a}^{2}) \bigg]
   \nonumber \\
 & &  \hspace{1cm} -\frac{9 \sin^{2}\theta}{k_{s\overline{s}}}
    \bigg[ \sum_{a}
   \gamma_{i}^{a} \left(\frac{g_{a}^{4}}{4}\right) \left(\ln(g_{a}^{2})
   +\frac{5}{3}\right)  \nonumber \\
 & &   \hspace{2.5cm}
+ \ln\big[(t +
   \overline{t})|\eta(t)|^4\big] \big( \sum_{a} \gamma_{i}^{a}
   \big(\frac{g_{a}^{4}}{4}\big) +\frac{1}{3} \sum_{jk} \gamma_{i}^{jk}
   k_{s} k_{\overline{s}} \big) \bigg] \Bigg\} \, , \\
A_{ijk} &=&  m_{3/2} \Bigg\{
  (\gamma_{i} + \gamma_{j} + \gamma_{k} )\bigg[ 1 + 2 t \cos\theta 
  G_{2} \bigg] \nonumber \\
 && \hspace*{1cm} +\frac{\sqrt{3} \sin\theta}{k_{s\overline{s}}^{1/2}}
\bigg[ k_{s} + \sum_{a} \frac{g_a^2}{2}
 ( \gamma_{i}^{a} +  \gamma_{j}^{a} +  \gamma_{k}^{a} )
(1 - \ln(g_a^2))  \nonumber \\
 & &  \hspace{2.8cm} -\ln\bigg[(t + \overline{t})
     |\eta(t)|^4\bigg]
  \big( \sum_{a} g_{a}^{2}( \gamma_{i}^{a} + \gamma_{j}^{a} + \gamma_{k}^{a} )
 \nonumber \\ && \hspace{5.8cm} -
  \sum_{lm} k_{s} (\gamma_{i}^{lm} + \gamma_{j}^{lm} + \gamma_{k}^{lm})
   \big) \bigg] \Bigg\} \, .
\label{eq:a3}
\end{eqnarray}
 In all three cases we have assumed that terms
proportional to the $\log(\tilde \mu_i)$ can be neglected
($\tilde \mu_i$ denote the Pauli Villar masses).

\subsubsection{General High Scale Model}

It is clear from the examples above that up to now there is no
unique mechanism for supersymmetry breaking. Therefore, we have implemented
the possibility to specify rather freely a high scale model. This model
is specified by: a set of three in principal non--universal gaugino mass 
parameters $M_{1/2}[U(1)]$,  $M_{1/2}[SU(2)]$,  $M_{1/2}[SU(3)]$;
a scalar mass for each type of sfermion, resulting in
fifteen parameters: $M^0_{\tilde E,ii}$, $ M^0_{\tilde L,ii}$, 
$ M^0_{\tilde D,ii}$,
$ M^0_{\tilde U,ii}$, $ M^0_{\tilde Q,ii}$; two Higgs mass parameters
$M^0_{H_1}$ and $M^0_{H_2}$; nine different $A$ parameters $A_{0,e,ii}$,
 $A_{0,d,ii}$ and  $A_{0,u,ii}$. Here $ii$ denotes that only the
diagonal entries can be set, because in the current version the
effects of generation mixing is not taken into account.
 A model of this kind has been used in \cite{Bartl:2001wc} 
for the study of low energy observables and the supersymmetric
spectrum. It also can be used, for example, to set the boundary conditions
for the gaugino mediated supersymmetry breaking \cite{Kaplan:1999ac}.
This general model will be denoted by SUGRA.

\subsubsection{General MSSM at low energies}
\label{subsec:MSSM}

Starting with version 2.2.0 there exists also the possibility 
to give the parameters at the low scale $M_{EWSB}$ together
with scale. In this case the parameters are taken as running
parameters at the scale $M_{EWSB}$ and the masses and mixing
angles are calculated using these parameters which in turn
serve as input for the calculation of decay widths and cross
sections. The input parameter are: the electroweak gaugino mass
parameters $M_1$ and $M_2$,  the gluino mass $m_{\tilde g}$;
the parameters describing the Higgs sector $\mu$, $\tan\beta$, the
mass of the pseudoscalar Higgs mass $m_A$; the
sfermion mass parameters  $M_{\tilde E,ii}$, $M_{\tilde L,ii}$, 
$M_{\tilde D,ii}$, $M_{\tilde U,ii}$, $M_{\tilde Q,ii}$, $A_{u,ii}$, 
$A_{d,ii}$, and $A_{l,ii}$.
This general model will be denoted by MSSM.

\section{Decays of supersymmetric particles and Higgs bosons}
\label{sect:decays}

The programs calculates the most important two- and three-body decays
of supersymmetric particles at tree level. 
 In case of three-body decays the formulas are
implemented such, that the effects of decay widths in the propagators
are taken into account \cite{ToBePublished}. 
Therefore, it is possible to perform the
calculation even in case that some of the intermediate particles are
on-shell. This is useful in the case that the two--body decays
have small phase space, because then the calculation of the three-body
decays gives a more accurate result, 
e.g.~$\Gamma(\tilde \chi^+_1 \to \tilde \chi^0_1 W^+) \times 
      \mathrm{BR}(W^+ \to \nu l^+)$ can be quite different from
$\Gamma(\tilde \chi^+_1 \to\tilde \chi^0_1 \nu l^+)$ if the 
decay $\tilde \chi^+_1 \to \tilde \chi^0_1 W^+$ has only small
phase space.

The following sfermion decays are calculated:
\begin{eqnarray}
 \label{eq:sfermiondecays}
 {\tilde f}_i &\to& f  \, {\tilde \chi}^0_k , \, \,
                    f'  \, {\tilde \chi}^\pm_l \\
 {\tilde f}_i  &\to& {\tilde f}_j  \, Z^0 , \, \,  {\tilde f}'_j  \, W^\pm \\
 { \tilde f}_i  &\to& {\tilde f}_j  \, (h^0, H^0, A^0), \, \,
                       {\tilde f}'_j  \, W^\pm
\end{eqnarray}
In case of the lighter stop, it is possible that all two-body decays
modes are kinematically 
forbidden at tree--level. In this case the following decay
modes are important \cite{Hikasa:1987db,Porod:1996at,djouadi3}:
\begin{eqnarray}
{\tilde t}_1 &\to& c \, {\tilde \chi}^0_{1,2} \\
{\tilde t}_1 &\to& W^+ \, b \, {\tilde \chi}^0_1 , \, \, 
                   H^+ \, b \, {\tilde \chi}^0_1 \\
{\tilde t}_1 &\to& b \, \nu \, {\tilde l}^+_i , \, \,
                   b \, l^+ \, {\tilde \nu}
\end{eqnarray}
where $l=e, \mu , \tau$. The corresponding widths are calculated within
\verb+SPheno+ using the formulas given in \cite{Porod:1996at}. In case of
GMSB models scenarios exist where the charged sleptons are next to
lightest supersymmetric particles (NLSP) and the gravitino $\tilde G$ is
the LSP. In this case the sleptons decay according to:
\begin{eqnarray}
 \label{eq:sleptondecays}
 {\tilde l}_i &\to& l  \, {\tilde G}
\end{eqnarray}
Here we use the formulas given in \cite{Giudice:1998bp}.

It is well known that the partial widths of sfermions can receive
considerable radiative corrections \cite{Kraml:1996kz}. However, the branching
ratios are not that strongly affected \cite{Bartl:2000kw}.
 Therefore, for the moment
being tree-level formulas are implemented. Some important numerical
effects of higher order
corrections are nevertheless implemented by using 1-loop corrected masses
and running couplings in the formulas.  The complete implementation of
higher-order corrections is left for future versions of the program.

In case of charginos and neutralinos the following decay modes are
calculated:
\begin{eqnarray}
 {\tilde \chi}^0_i &\to& Z^0  \, {\tilde \chi}^0_j , \, \,
                         W^\pm     \, {\tilde \chi}^\mp_k \\
 {\tilde \chi}^0_i &\to& (h^0, H^0, A^0)  \, {\tilde \chi}^0_j , \, \,
                         H^\pm     \, {\tilde \chi}^\mp_k \\
 {\tilde \chi}^0_i &\to& f \bar{\tilde{f}_j}, \, \,  \bar{f} \tilde{f}_j \\
 {\tilde \chi}^+_k&\to&  Z^0  \, {\tilde \chi}^+_s , \, \,
                         W^+     \, {\tilde \chi}^0_j \\
 {\tilde \chi}^+_k&\to&  (h^0, H^0, A^0)   \, {\tilde \chi}^+_s , \, \,
                         H^+     \, {\tilde \chi}^0_j \\
  {\tilde \chi}^+_k&\to&  f {\tilde f}_i'
\end{eqnarray}
In case that all two body decay modes are kinematically forbidden the following
three--body decays are calculated:
\begin{eqnarray}
 {\tilde \chi}^0_i &\to& f \, \bar{f}  \, {\tilde \chi}^0_j , \, \,
                         f \, f'     \, {\tilde \chi}^\mp_k \\
 {\tilde \chi}^0_i &\to& q \, \bar{q}  \, {\tilde g}  \\
 {\tilde \chi}^+_k &\to& f \, \bar{f}  \, {\tilde \chi}^+_s , \, \,
                         f \, f'     \, {\tilde \chi}^0_j \\
 {\tilde \chi}^+_k &\to& q \, q'     \, {\tilde g}
\end{eqnarray}
In the calculation we have included all contributions from gauge bosons,
sfermions and Higgs  bosons \cite{ToBePublished,Baer:1998bj}. 
The Higgs contributions can be important in certain
regions of parameter space \cite{Bartl:1999iw}. In addition the loop
induced decay s
\begin{eqnarray}
{\tilde \chi}^0_i &\to& {\tilde \chi}^0_j \, \gamma
\end{eqnarray}
are calculated \cite{Haber:1988px} taking into account the left-right mixing
of sfermions.
Similarly to case
of the sleptons there exist parameter regions in GMSB models where
the lightest neutralino is the NLSP and it decays according to
\begin{eqnarray}
 {\tilde \chi}^0_1 &\to& \gamma \, \ {\tilde G} \\
 {\tilde \chi}^0_1 &\to& Z^0 \, \ {\tilde G} \\
 {\tilde \chi}^0_1 &\to& h^0 \, \ {\tilde G} 
\end{eqnarray}
Here we use the formulas given in \cite{Giudice:1998bp}.

In case of gluinos the following two--body decays are
calculated:
\begin{eqnarray}
 \tilde g \to q \, {\tilde q}_i
\end{eqnarray}
with $q=u, d, c, s, t, b$.
Again, in case that these decays are kinematically suppressed, the
three-body decay modes are calculated:
\begin{eqnarray}
\tilde g &\to& \tilde \chi^0_i \, q \, \bar{q} \\
\tilde g &\to& \tilde \chi^\pm_j \, q' \, \bar{q} \\
\tilde g &\to& \bar{b} \, W^- \, \tilde t_1,  b \, W^+ \, \tilde t_1^*
\end{eqnarray}
Here we have implemented the formulas given in
\cite{glu1}. In addition the decays
\begin{eqnarray}
{\tilde g} &\to& {\tilde \chi}^0_i \, \gamma
\end{eqnarray}
are calculated \cite{Haber:1988px,Bartl:1990ay} 
taking into account the left-right mixing
of sfermions.

In case of Higgs bosons the following decays are calculated:
\begin{eqnarray}
 \phi &\to& f \, \bar{f} \\
 \phi &\to&  \tilde f_i \, \bar{\tilde f}_j \\
 \phi &\to& {\tilde \chi}^0_k \, {\tilde \chi}^0_l \\
 \phi &\to& {\tilde \chi}^+_r \, {\tilde \chi}^-_s \\
 H^0 &\to& Z^0 \, Z^0 , \, \,  W^+ \, W^-\\
 H^0 &\to& h^0 \, h^0 \\
 A^0 &\to& h^0 \, Z^0 \\
 H^+ &\to& f \, \bar{f}' \\
 H^+  &\to&  \tilde f_i \, \bar{\tilde f}_j' \\
 H^+ &\to& {\tilde \chi}^0_k \, {\tilde \chi}^-_s \\
 H^+ &\to& h^0  \, W^+
\end{eqnarray}
with $\phi=h^0, H^0, A^0$ and $f=\nu_i, e, \mu, \tau, u, d, c, s, t, b$.
It is well known, that the widths as well as the branching ratios of the 
Higgs bosons can receive large one--loop corrections 
\cite{Drees:1990dq,Eberl:2001vb,Djouadi:mr}.
In the present version only the gluonic QCD corrections for the decays into
quarks \cite{Drees:1990dq} have been implemented.
Therefore, the numbers provided by \verb+SPheno+ have to be taken with care
and for refined analysis other programs, such as {\sf HDECAY}
\cite{Djouadi:1997yw} should be used.

\section{Production of supersymmetric particles and Higgs bosons}
\label{sect:prod}

The program calculates the following cross sections:
\begin{eqnarray}
 e^+ \, e^- \, &\to& \tilde f_i \, \tilde f_j \hspace{1cm} (f=l, \nu, q) \\
 e^+ \, e^- \, &\to& \tilde \chi^0_k \, \tilde \chi^0_n \\
 e^+ \, e^- \, &\to& \tilde \chi^+_r \, \tilde \chi^-_s \\
  e^+ \, e^- \, &\to& h^0 \, Z , \, \, H^0 Z \\
  e^+ \, e^- \, &\to& h^0 \, A^0 , \, \, H^0 A^0 \\
  e^+ \, e^- \, &\to&  H^+ H^-
\end{eqnarray}
We haven taken the formulas of \cite{Bartl:1997yi} for sfermion
production, \cite{Moortgat-Pick:2000uz,ToBePublished} for production of
charginos and
neutralinos and \cite{Djouadi:mr} for Higgs boson production.
Initial state radiation has been included using the formula given in 
\cite{drees1}. In case
of squarks in addition QCD corrections due to gluon exchange are included
\cite{drees1,Eberl:1996wa}. Care has to be taken in case one calculates
the cross sections near threshold because then higher order corrections
are important to get reliable results \cite{Freitas:2001zh} and, thus, the
numbers obtained in the program have to be taken with care near the
threshold.
  All cross
sections are implemented such, that one can specify the degree of
longitudinal polarization $P_{e^-}$ of the incoming electron beam 
as well as the degree of longitudinal polarization  $P_{e^+}$ of the 
incoming positron beam.
Here $P_{e^-}$ is within the range $[-1,1]$, where $\{-1,0,1\}$
denote 100\% left-handed electrons, completely unpolarized electrons
and 100\% right-handed electrons, respectively. 
The same notation is used in case of
positrons. For example, $P_{e^-}=-0.8$ ($P_{e^+}=-0.8$) means that
80\% of the electrons (positrons) are left-polarized whereas the
remaining 20\% are unpolarized.

\section{Low Energy Constraints}
\label{sec:constraints}

The supersymmetric parameters are constrained by direct searches at
colliders and by loop-effects which supersymmetric particles induce
observables of  low energy
experiments. Provided one neglects mixing between different sfermion
generations the following quantities constrain several parameters of
the MSSM: the rare decay $b \to s \, \gamma$, the anomalous magnetic
moment of the muon $a_\mu$ and the supersymmetric contributions to the $\rho$
parameter. These constraints are implemented in the program using the
formulas given in \cite{Bertolini:1990if,Cho:1996we} for $b \to s \, \gamma$
supplemented by the QCD corrections as given in \cite{kagan},
\cite{Ibrahim:1999hh} for $a_\mu$ and \cite{Drees90} for the sfermion
contributions to the $\rho$ parameter. In call cases we use the
running couplings at $m_Z$ for the calculation of the observables.
The use of running couplings together with the correct implementation
of supersymmetric threshold corrections for the couplings
results in taking into account
the most important higher oder corrections as has been pointed out e.g.~in
\cite{Carena:2000uj,Buras:2002vd} for the case of $b \to s \, \gamma$.  The
implementation of the supersymmetric threshold corrections to the
couplings will be
discussed in the next section.

\section{Details of the Calculation}
\label{sec:calc}

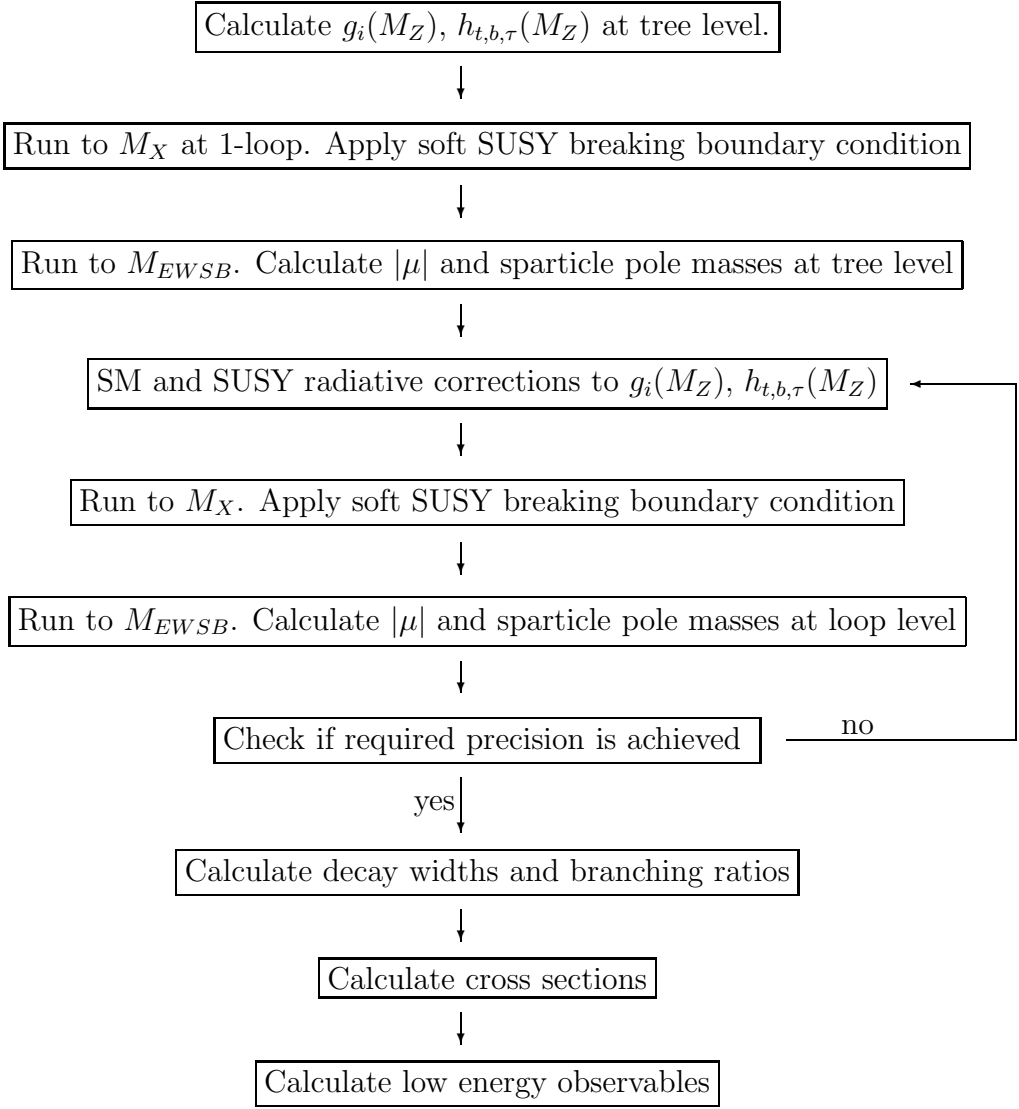
\begin{figure}
\begin{center}
\begin{picture}(350,450)
\put(40,410){\makebox(280,10)[c]{\fbox{Calculate
                                    $g_i(M_Z)$, $h_{t,b,\tau}(M_Z)$ 
                                     at tree level.}}}
\put(170,400){\vector(0,-1){12}}
\put(40,365){\makebox(280,10)[c]{\fbox{Run to $M_X$ at 1-loop.
                      Apply soft SUSY breaking boundary condition}}}
\put(170,355){\vector(0,-1){12}}
\put(40,320){\makebox(280,10)[c]{\fbox{Run to $M_{EWSB}$. Calculate $|\mu|$ and
                      sparticle pole masses at tree level}}}
\put(170,310){\vector(0,-1){12}}
\put(40,275){\makebox(280,10)[c]{\fbox{SM and SUSY radiative corrections to
                                    $g_i(M_Z)$, $h_{t,b,\tau}(M_Z)$}}}
\put(170,265){\vector(0,-1){12}}
\put(40,230){\makebox(280,10)[c]{\fbox{Run to $M_X$. Apply soft SUSY breaking
                                      boundary condition}}}
\put(170,220){\vector(0,-1){12}}
\put(40,185){\makebox(280,10)[c]{\fbox{Run to $M_{EWSB}$. Calculate $|\mu|$ and
                      sparticle pole masses at loop level}}}
\put(170,175){\vector(0,-1){12}}
\put(40,140){\makebox(280,10)[c]{\fbox{Check if required precision is achieved
            }}}
\put(20,115){\makebox(280,10)[c]{yes}}
\put(170,131){\vector(0,-1){21}}
\put(180,145){\makebox(280,10)[c]{no}}
\put(293,145){\line(1,0){87}}
\put(380,145){\line(0,1){135}}
\put(380,280){\vector(-1,0){40}}
\put(40,90){\makebox(280,10)[c]{\fbox{Calculate decay widths
            and branching ratios}}}
\put(170,81){\vector(0,-1){12}}
\put(40,50){\makebox(280,10)[c]{\fbox{Calculate cross sections}}}
\put(170,42){\vector(0,-1){12}}
\put(40,10){\makebox(280,10)[c]{\fbox{Calculate low energy observables}}}
\end{picture}
\end{center}
\caption{Algorithm used to calculate the SUSY spectrum, decay widths,
production cross sections and low energy observables. Each step
(represented by a box) is explained in the text. The initial step is the
uppermost one. $M_{EWSB}$ is the scale at which the EWSB
conditions
are imposed, and $M_X$ is the scale at which the high
energy SUSY breaking boundary conditions are imposed.}
\label{fig:algorithm}
\end{figure}

In this section we describe the algorithm used. It is
schematically displayed in \fig{fig:algorithm}. The following
standard model parameters are used as input: fermion masses, the $Z$-boson pole
mass, the fine structure constant $\alpha$, the Fermi constant $G_F$
and the strong coupling constant $\alpha_s(m_Z)$. It is assumed that
  $\alpha_s(m_Z)$ is given in the $\msbar$ scheme. We describe first
the implementation of the high scale models and comment then on the
case of the implemented MSSM model.

\subsection{First rough calculation of SUSY and Higgs boson masses}

In a first step, we calculate gauge and Yukawa couplings at $m_Z$
scale using tree-level formulas. These are used as input for the one--loop
RGEs to get the gauge and Yukawa couplings at the high scale where
also the boundary conditions for the high scale model under study are imposed. 
Afterward
one--loop RGEs are used to get a first set of parameters at the electroweak
scale. These parameters are used to get a first set of supersymmetric particle
masses and Higgs masses using tree-level formulas except for the neutral
CP-even Higgs bosons where one-loop effects due to (s)quarks of the third
generation are taken into account. 
These masses are the starting point for
the iterative loop which calculates the spectrum within the required 
precision as described below.

\subsection{Main loop for the calculation of SUSY and Higgs boson masses}

In the next step the gauge couplings and $\sin^2 \theta_W$ 
are calculated at $m^2_Z$ in the $\drbar$ scheme using the formulas given
Appendix C of \cite{Pierce:1996zz}.  The only difference is that we use an 
updated
number for $\Delta \alpha_{lep} + \Delta \alpha_{had} = 0.06907$ 
\cite{Jegerlehner:2003qp}.
The mass of the $W$-boson is calculated using the
formula \cite{Degrassi:1991tu}
\begin{eqnarray}
 m_W^2 &=& m_Z^2  \hat \rho  \left( \frac{1}{2} 
       + \sqrt{ \frac{1}{4} - \frac{\alpha^{\drbar}(m_Z) \pi}
            {\sqrt{2} G_F m_Z^2 \hat \rho (1 - \Delta \hat r)}}
  \right)
 \label{eq:mw-onshell} \\
\hat \rho &=& \frac{1}{1 - \Delta \hat \rho} \\
\Delta \hat \rho &=& Re \left( \frac{ \Pi^T_{ZZ}(m_Z^2)}{\hat \rho m_Z^2}
        - \frac{ \Pi^T_{WW}(m_W^2)}{m_W^2} \right) \\
\Delta \hat r &=& \hat \rho \frac{ \Pi^T_{WW}(0)}{m_W^2} 
   -  \frac{ \Pi^T_{ZZ}(m_Z^2)}{\hat m_Z^2} + \delta_{VB} 
\end{eqnarray}
where $\Pi^T_{VV}$ is the transverse part of the vector boson self-energy
and $\delta_{VB}$ contains non-universal corrections to the $\mu$ decay.
In the computation we have included the leading SM two-loop corrections
the formulas of \cite{Fanchiotti:1993tu} and adapting the Higgs contribution as
 in \cite{Pierce:1996zz}. The formulas of the SUSY contribution to
$\delta_{VB}$ are given in \cite{Grifols:1985xs,Chankowski:1994eu}. 
All masses appearing
in the loops are running except for the top-quark, because the
2-loop part is given for an on-shell definition of the top mass. Note here
that also the gauge boson masses in the loops are running masses and thus
an iteration has to be performed in practice.

For the calculation of the 
Yukawa couplings we use the complete formulas for the fermion masses
and the vacuum expectation values given in Appendix D of
\cite{Pierce:1996zz}. In addition we have implemented the following 
improvements.
The five light quarks and $\alpha_s$ 
are evolved to $m_Z$ using 4-loop RGEs as given
in \cite{vanRitbergen:1997va} including 
threshold corrections at the low scale 
\cite{Chetyrkin:1997sg}.
This evolution is done in the $\msbar$ scheme. At $m_Z$ the shift to
the $\drbar$ scheme is performed using the formulas given in 
\cite{Avdeev:1997sz,Baer:2002ek}, e.g.:
\begin{eqnarray}
m_{b',{\drbar}}(m_Z) &=& m_{b,\msbar}(m_Z)
  \left( 1 - \frac{\alpha_s}{3 \pi}
       - \frac{23 \alpha_s^2}{72 \pi^2}
       + \frac{4 g^2_2}{128 \pi^2} - \frac{13 {g'}^2}{1152 \pi^2} \right)
\end{eqnarray}
where $\alpha_s$ is given in the $\drbar$ scheme which is the reason for
the different factor in front of  $\alpha_s^2$ compared to \cite{Baer:2002ek}.
We use the complete formulas given in  Appendix D of \cite{Pierce:1996zz}
to calculate the SUSY contribution denoted by $\Delta m_{b,SUSY}(m_Z)$.
For the calculation we use running gauge and Yukawa couplings at $m_Z$.
The SUSY contributions $\Delta m_{b,SUSY}(m_Z)$
are then resumed using \cite{Carena:1999py}
\begin{eqnarray}
m_{b,\drbar}(m_Z) = \frac{m_{b',\drbar}(m_Z)}
                         {1 - \frac{\Delta m_{b,SUSY}(m_Z)}{m_{b,\drbar}(m_Z)}}
\label{eq:resumb}
\end{eqnarray}

In the case of the top-quark a modified procedure has to be used due to
its large mass. In this case we start from the pole mass $m_t$ and use
the formulas given in \cite{Pierce:1996zz} to obtain $m_t^{\drbar}(m_Z)$.
The difference compared to \cite{Pierce:1996zz} is in the 2-loop parts
where we have taken the exact 2-loop formula given in 
\cite{Avdeev:1997sz}:
\begin{eqnarray}
\Sigma_t^{2l} &=& - \left(\frac{\alpha_s(m_Z)}{\pi}\right)^2
  \left( \frac{8 \pi^2}{9} + \frac{2011}{18} + \frac{16 \pi^2 \ln 2}{9}
       - \frac{8 \zeta(3)}{3}
      + \frac{246}{3} L + 22 L^2   \right) \\
L &=& \log\left(\frac{m^2_Z}{m^2_t}\right)
\end{eqnarray}

In the case of leptons we first calculate the running masses $m_l(m_l)$
at 1-loop level in the $\msbar$ scheme. 
Afterward we use two-loop RGEs as given in 
\cite{Arason:1992ic} to evolve them to $m_Z$. Here we perform the shift to
the $\drbar$ scheme using the formula:
\begin{eqnarray}
m_l^{\drbar,SM}(m_Z) &=& m_l^{\msbar,SM}(m_Z)
  \left(1 - \frac{3 ({g'}^2-g_2^2)}{128 \pi^2} \right)
\end{eqnarray}
The analog of \eq{eq:resumb} is used to obtain the running mass including
the SUSY threshold corrections.

The obtained gauge and Yukawa couplings are evolved to the high scale
using two--loop RGEs \cite{Martin:1993zk}. The high scale can either
be fixed or can be calculated from the requirement  $g_1 = g_2$ at the 
high scale. The various
cases are discussed in Sect.~\ref{sec:example} and
Appendix~\ref{app:switch}. At the high scale the boundary
conditions for the soft SUSY breaking parameters are set. The implemented
models are summarized in Sect.~\ref{sect:models}; see also 
Appendix~\ref{app:model}.
The complete
set of two--loop RGEs \cite{Martin:1993zk} is used to evolve the
parameters down to the electroweak scale 
$M_{EWSB} = \sqrt{m_{\tilde t_1} m_{\tilde t_2}}$ (there exists also the
possibility to fix $M_{EWSB}$ to a constant value as described in
Appendix~\ref{app:switch}). 

The parameters are used as input to calculate the sparticle pole
masses at one--loop order and in case of the neutral Higgs at
two--loop order.  Here we use the complete formulae given in the
appendices of \cite{Pierce:1996zz} for the one--loop contributions and
for the 2-loop corrections 
$O(\alpha_s (\alpha_t + \alpha_b)+ (\alpha_t + \alpha_b)^2 +
\alpha_\tau \alpha_b + \alpha^2_tau)$ 
for the neutral Higgs boson the formulas given in
\cite{Degrassi:2001yf,Dedes:2002dy,ADKPS}.
  For the $O(\alpha_s \alpha_b)$ contributions
we use the complete expressions which can be obtained from the
$O(\alpha_s \alpha_t)$ contributions by appropriate replacements.  In
case of sfermions we have included in all cases left--right mixing.
All gauge and Yukawa couplings are understood as $\drbar$ quantities
at $M_{EWSB}$. Also $\tan\beta$ and the vacuum expectation values are
evaluated at $M_{EWSB}$ to get a consistent set of input parameters.
Note, that we express in all couplings the fermion masses and gauge
boson masses by their corresponding expressions due to
gauge couplings, Yukawa couplings and vacuum expectations values in
the formulas of \cite{Pierce:1996zz}. In all cases running masses are
used as input for the loop integrals.  In addition we
have implemented the 
$O(\alpha_s (\alpha_t + \alpha_b)+ (\alpha_t + \alpha_b)^2 +
\alpha_\tau \alpha_b + \alpha^2_tau)$ 
corrections for the calculation of $|\mu|$
\cite{Dedes:2002dy,ADKPS}. The numerical evolution
of the one--loop integrals is based on the FF package 
\cite{vanOldenborgh:1989wn} and the
LoopTools package \cite{Hahn:1998yk}.

The masses obtained are used as input for the next iteration which starts
again by calculating the SUSY contributions to gauge and Yukawa couplings
at $m_Z$. In the case that during this iterative
process an unphysical situation occurs, e.g.~a pole mass squared being
negative, the program terminates and it provides information on the exact
reason for termination.  
 The iteration is continued until all relative differences between
the sparticle masses are smaller then the user imposed quantity $\delta$:
\begin{eqnarray}
 \delta > \frac{|m_i - m_{i-1}|}{m_i} 
\end{eqnarray}
for all sparticle masses; i denotes the i-th iteration.
In most cases this achieved after three
to four iterations. 
In the case that more than the maximal allowed number of iterations
(user specified) are
necessary, the program leaves the iteration giving a warning message.

In the case of the MSSM model the running between $M_{EWSB}$ and
$M_X$ is obviously omitted. In this model one has, however, to preform
an iteration to get the Yukawa and gauge couplings correct because
$\tan\beta$ is defined at $M_{EWSB}$ instead of $m_Z$.

\subsection{Calculation of the other observables}

The masses and mixing angles are then used to calculate the branching
ratios and decay widths. Here two- and three body decays of supersymmetric
particles are calculate. Note that we use the couplings as input which 
are renormalized at the scale $M_{EWSB}$.
The user has the possibility to force the
program to calculate three body decays even if one or more of the
intermediate particles are on-shell. This possibility is useful in the
case where the 2-body decay  has only small phase because then the
calculation of the three body decay width(s) give a more reliable
result.

Afterward the  production cross sections at an $e^+ e^-$ collider of
all supersymmetric particles as well as all Higgs bosons are
calculated. Here the user has the possibility to specify the center 
of mass energy as well as the degree of longitudinal polarization of
the incoming beams. Moreover, the user can specify if initial state
radiation shall be included in the calculation or not.

Finally,
the low--energy constraints described in Section~\ref{sec:constraints}
are calculated: $BR(b \to s \gamma)$, 
SUSY contributions to $a_\mu$
and the sfermion contributions to $\Delta \rho$. For the the calculation 
of these quantities we evolve the gauge and Yukawa couplings from the scale 
$M_{EWSB}$ down to $m_Z$. The couplings at $m_Z$ are then used as input
for the calculation of the low energy observables.
For example in calculation of
$BR(b \to s \gamma)$ the most important contributions to 
the $C_7$ coefficients are
implemented as
\begin{eqnarray}
 C_7(W^+) &=& -\frac{K_{ts} K_{tb}x_{tW}}{4 \, m^2_W}  
\left(\frac{2}{3} F_1(x_{tW})+ F_2(x_{tW}) \right) \\
 C_7(H^+) &=& - \frac{K_{ts} K_{tb}}{4 \,m^2_{H^+} }
   \left[ \frac{Y_t^2 \cos^2\beta}{4}
   \left(\frac{2}{3} F_1(x_{tH^+})+ F_2(x_{tH^+}) \right)
  \right. \nonumber \\ &&\hspace{1.6cm} \left. 
    -  \frac{Y_b Y_t \cos \beta \sin \beta m_t}{m_b}
   \left(\frac{2}{3} F_3(x_{tH^+})+ F_4(x_{tH^+}) \right)
  \right]  \\
 C_7({\tilde \chi}^+) &=& \sum_{i,j=1}^2 \frac{K_{ts} K_{tb}}
                                              {4 \, m^2_{\tilde t_i}}
   \left[ C^2_{R,ij}
   \left(\frac{2}{3} F_2(x_{{\tilde \chi}^+_j \tilde t_i})
   + F_1(x_{{\tilde \chi}^+_j \tilde t_i}) \right)
  \right. \nonumber \\ &&\hspace{1.6cm} \left. 
    - C_{L,ij} C_{R,ij}
   \left(\frac{2}{3} F_4(x_{{\tilde \chi}^+_j \tilde t_i})
        + F_3(x_{{\tilde \chi}^+_j \tilde t_i}) \right)
  \right]  \\
C_{L,ij} &=& Y_b R_{\tilde t,i1} U_{j2} \\
C_{R,ij} &=& -g  R_{\tilde t,i1} V_{j1} +  Y_t R_{\tilde t,i2} V_{j2}
\end{eqnarray}
Here $Y_i$ are the Yukawa couplings, $U$ and $V$ are the
chargino matrices, $K$ is the CKM matrix,
$R_{\tilde t}$ is the stop mixing matrix 
and $x_{ab} = m^2_a/m^2_b$.
The loop functions $F_i$ are given in \cite{Bertolini:1990if}.
A similar replacement
is done in the contributions to the $C_8$ coefficient.
Moreover, in the program also the contributions from the first two generation
of (s)fermions to $C_{7,8}$ are included for completeness.
We then use \cite{kagan} to obtain
\begin{eqnarray}
BR(b \to s \gamma) = 1.258 + 0.382  r7^2 + 0.015  r_8^2 
       + 1.395 r_7 + 0.161 r_8
        + 0.083 r_7 r_8
\end{eqnarray}
where $r_7 = C_7/C_7(W^+)$ and  $r_8 = C_8/C_8(W^+)$. 
In this way  important higher order corrections are
taken into account, in particular the large $\tan\beta$ effects in
case of $b \to s \gamma$ \cite{Buras:2002vd}.

\section{A sample example}
\label{sec:example}

In this section we discuss the executable statements of the main program
given in the file \verb+SPheno.f90+. 
In the first statements the required modules are loaded and the various
variables are defined.
Afterward the error system is initialized and the input data
are read by calling:
\begin{verbatim}
 Call ReadingData(kont)
\end{verbatim}
The source code can be found in the file \verb+SPheno.f90+.
This routine checks first if the file \verb+LesHouches.in+ exists where
the input data are provided according to SLHA
(for a detailed description see Appendix~\ref{app:SLHA}). 
If this file is absent,
the routine checks if the files
\verb+Control.in+ and/or \verb+StandardModel.in+ are present. The first one
can be used to set various flags (see Appendix~\ref{app:Control}) whereas
the second one is used to set the SM input (see Appendix~\ref{app:SM}).
Standard values as described in the Appendix are used if any of these files
is absent. Afterward the model specific information is
read from the file \verb+HighScale.in+ which is described in 
Appendix~\ref{app:model} (for a short description of the implemented models
see \sect{sect:models}).
In all cases the file  \verb+Messages.out+ is opened 
at channel 10 where
all warnings and/or debugging informations are stored.

 Before calling
the  subroutine \verb+CalculateSpectrum+, which performs the calculation
of the spectrum, 
the user has the possibility to fix 
the high scale and/or the scale where the parameters and the
loop corrected masses are calculated. For this purpose one or both
of the following lines must be uncommented in the program:
\begin{verbatim}
! Call SetGUTScale(2.e16_dp)   ! please put the GUT scale
! Call SetRGEScale(1.e3_dp**2) ! please put the scale M_EWSB squared
\end{verbatim}
The default is that these scales are calculated by the program. The high
scale is computed from the requirement $g_1=g_2$ (except in GMSB where
the high scale is an input). The scale $M_{EWSB}$ is given by
$M_{EWSB}=\sqrt{m_{\tilde t_1} m_{\tilde t_2}}$.

\begin{table}
\caption{Variables for parameters and couplings.
  The parameters are defined in the module {\tt MSSM\_Data} and they are
  explained in Section~\ref{sect:MSSM}. dp means double
  precision.}
\label{tab:parameters}
\begin{tabular}{ll}
parameter/coupling & type \& Fortran name  \\
$e^{\varphi_\mu}$ & \verb+complex(dp) :: phase_mu+ \\
$\tan \beta$  & \verb+real(dp) :: tanb+     \\
$M_1, M_2, M_3$ & \verb+complex(dp) :: Mi(3)+ \\
$M^2_E$, $M^2_L$ & \verb+complex(dp), dimension(3,3) :: M2_E, M2_L+ \\
$M^2_D$, $M^2_Q$, $M^2_U$      
  & \verb+complex(dp), dimension(3,3) :: M2_D, M2_Q, M2_U+ \\
$A_l, A_d, A_u$& \verb+complex(dp), dimension(3,3) :: A_l, A_d, A_u+ \\
$\mu$    &\verb+complex(dp) :: mu+ \\
$B \mu$    &\verb+complex(dp) :: B+ \\
$M^2_H$       & \verb+real(dp) :: M2_H(2)+ \\
$g'$, $g$   & \verb+real(dp) :: gp, g+ \\
$Y_l, Y_d, Y_u$& \verb+complex(dp), dimension(3,3) :: Y_l, Y_d, Y_u+ \\
$v_1, v_2$ & \verb+real(dp) :: vevSM(2)+ \\
$g'$, $g$, $g_s$   & \verb+real(dp) :: gauge(3)+ \\
\end{tabular}
\end{table}

\begin{table}[t]
\caption{Variables for masses and mixing matrices as given by the routine
         {\tt CalculateSpectrum}. They are defined in the module {\tt MSSM\_Data}
         and their connection to the parameters at tree--level 
         is explained explained in Section~\ref{sect:MSSM}. dp means
         double precision.}
\label{tab:masses}
\begin{tabular}{ll}
masses / mixing matrix & type \& Fortran name  \\
$m_{\tilde g}$ & real(dp) :: mglu \\
$e^{\varphi_{\tilde g}}$ & complex(dp) :: PhaseGlu \\
$m_{\tilde \chi^+_i}$ &  real(dp) :: mC(2)  \\
$U$, $V$ &  complex(dp) :: U(2,2), V(2,2)    \\
$m_{\tilde \chi^0_j}$ &  real(dp) :: N(4) \\
$N$ &  complex(dp) :: N(4,4)    \\
$m_{h^0}, m_{H^0}$ & real(dp) :: mS0(2) \\
$R_\alpha$ &  real(dp) :: RS0(2,2)    \\
$m_{G^0}, m_{A^0}$ &  real(dp) :: mP0(2) \\
$R_\beta$ &  real(dp) :: RP0(2,2)    \\
$m_{G^+}, m_{H^+}$ &  real(dp) :: mSpm(2) \\
$R'_\beta$ &  complex(dp) :: RSpm(2,2)    \\
$m_{\tilde \nu}$ & real(dp) :: mSneut(3) \\
$R_{\tilde \nu}$ & complex(dp) :: Rsneut(3,3) \\
$m_{\tilde l}$ & real(dp) :: mSlepton(6) \\
$R_{\tilde l}$ & complex(dp) :: Rslepton(6,6) \\
$m_{\tilde u}$ & real(dp) :: mSup(6) \\
$R_{\tilde u}$ & complex(dp) :: Rsup(6,6) \\
$m_{\tilde d}$ & real(dp) :: mSdown(6) \\
$R_{\tilde d}$ & complex(dp) :: Rsdown(6,6) \\
\end{tabular}
\end{table}

The accurate calculation of the SUSY parameters and the spectrum is done
by the following call:
\begin{verbatim}
 delta = 1.e-3_dp
 WriteOut = .False.
 n_run = 20
 If (kont.Eq.0) Call CalculateSpectrum(n_run, delta, WriteOut, kont, tanb, vevSM &
       & , mC, U, V, mN, N, mS0, mS02, RS0, mP0, mP02, RP0, mSpm, mSpm2, RSpm    &
       & , mSdown, mSdown2, RSdown, mSup, mSup2, RSup, mSlepton, mSlepton2       &
       & , RSlepton, mSneut, mSneut2, RSneut, mGlu, PhaseGlu, gauge, Y_l, Y_d    &
       & , Y_u, Mi, A_l, A_d, A_u, M2_E, M2_L, M2_D, M2_Q, M2_U, M2_H, mu, B     &
       & , m_GUT)
\end{verbatim}
The meaning of the various variables and their type is given in 
Tables~\ref{tab:parameters} and \ref{tab:masses}. 
Variable names ending with ``2'' indicate
masses squared. The variables for the mixing matrices are already
structured for a latter extension to include the effects of generation
mixing and/or complex phases:  the
sfermion mixing matrices are $6\times 6$ (except for sneutrinos which is
a $3\times 3$ matrix). In the present release most of the entries
are zero except for the diagonal $2\times 2$ blocks which contain the
left--right mixing for every species of sfermions. For example, the
$11$, $12$, $21$ and $22$ entries in \verb+Rslepton+ specify the left--right
mixing of selectrons, and similarly the $33$, $34$, $43$ and $44$ 
($55$, $56$, $65$ and $66$) entries specify the left--right
mixing of smuons (staus). The squark mixing matrices have the same
generation structure. This structure has been chosen to facilitate 
a later extension which includes flavour violating entries.
Beside the variables given in Tables~\ref{tab:parameters} and \ref{tab:masses}
 the following  variables appear:
\begin{itemize}
 \item \verb+delta+ : specifies the required relative precision on the masses.
   If the maximal relative differences between
  the physical masses obtained between two runs is smaller than  \verb+delta+,
   the routine \verb+Sugra+ leaves the iteration loop.
 \item \verb+m_GUT+ : the value of the scale where the high energy boundary
     conditions are imposed.
 \item \verb+kont+ : A variable which is 0 provided everything is o.k.
       Otherwise
     either a numerical problem has occurred and/or the parameters belong
     to an unphysical region, e.g.~a minimum of the potential where
     charge and/or colour breaking minima occur. In such a case the information
     is written to the file \verb+Messages.out+.
 \item \verb+WriteOut+ : If it is set \verb+.True.+ then 
     intermediate debugging information is written to the screen and the
     file \verb+Messages.out+. 
 \item \verb+n_run+ specifies the maximal number of iterations of the  main
     loop. A warning will be given in the case that the required precision
      \verb+delta+ has not been reached within \verb+n_run+  iterations.
\end{itemize}
Note that the parameters are running parameters at the scale $M_{EWSB}$.
The complete spectrum is calculated at 1--loop level using the formulas given
in \cite{Pierce:1996zz}. The exceptions
are the masses of the neutral Higgs bosons (scalar and pseudo-scalar) as
well as $\mu$ the two loop corrections due to $\alpha_s$ and all third
generation Yukawa couplings are included \cite{Degrassi:2001yf,Dedes:2002dy}.

In the next part the branching ratios, the partial decay widths and the
total  decay widths are calculated provided that \verb+L_BR=.TRUE.+ and
\verb+kont.eq.0+:
\begin{verbatim}
 If ((L_BR).and.(kont.eq.0)) then
  epsI = 1.e-5_dp
  deltaM = 1.e-3_dp 
  CalcTBD = .False.
  ratioWoM = 1.e-4_dp
  If (HighScaleModel.Eq."GMSB")  Is_GMSB = .True.
  Call CalculateBR(gauge, mGlu, PhaseGlu, mC, U, V, mN, N   &
     & , mSneut, RSneut, mSlepton, RSlepton, mSup, RSup     &
     & , mSdown, RSdown, uL_L, uL_R, uD_L, uD_R, uU_L, uU_R &
     & , mS0, RS0, mP0, RP0, mSpm, RSpm, epsI, deltaM       &
     & , CalcTBD, kont, ratioWoM, Y_d, A_d, Y_l, A_l, Y_u   &
     & , A_u, mu, vevSM, Fgmsb, m32                         &
     & , gP_Sl, gT_Sl, BR_Sl, gP_Sn, gT_Sn, BR_Sn, gP_Sd    &
     & , gT_Sd, BR_Sd, gP_Su, gT_Su, BR_Su, gP_C, gT_C      &
     & , BR_C, gP_N, gT_N, BR_N, gP_G, gT_G, BR_G, gP_P0    &
     & , gT_P0, BR_P0, gP_S0, gT_S0, BR_S0, gP_Spm, gT_Spm  &
     & , BR_Spm)
  end if
\end{verbatim}
Variables starting with \verb+gP_+, \verb+gT_+ and
\verb+BR_+ indicate partial widths, total widths and branching ratios,
respectively; they are \verb+Real(dp)+ vectors. The first index
is the index of the decaying particle whereas the second one
gives the mode.
 The correspondence between the second index and the modes
 is summarized for sfermions (variables ending
\verb+Sl+, \verb+Sn+, \verb+Sd+ and \verb+Su+ for sleptons,
sneutrino, d-squarks and u-squarks, respectively)
 in Table~\ref{tab:sfermiondecays}, for charginos in 
Table~\ref{tab:charginodecays}, for neutralinos in 
Table~\ref{tab:neutralinodecays}, for gluinos in Table~\ref{tab:gluinodecays}
and for the Higgs bosons in Tables~\ref{tab:higgsdecays} and 
\ref{tab:Hpdecays}.

\begin{table}[t]
\caption{Correspondence between the indices for sfermion partial widths
(branching ratios) and the modes. $\tilde t_1$ is listed extra because
it can have three-body decay modes. }
\label{tab:sfermiondecays}
\begin{center}
\begin{tabular}{l|c|c|c|c|c} 
mode & $\tilde l$ & $\tilde \nu$ & $\tilde d$ & $\tilde u$ & $\tilde t_1$\\
\hline
 $\tilde f_i \to f {\tilde \chi}^0_k$ & 1-4 & 1-4 & 1-4 & 1-4  & 1-4\\ 
 $\tilde f_i \to f' {\tilde \chi}^\pm_j$ & 5-6 & 5-6 & 5-6 & 5-6 & 5-6\\ 
 $\tilde f_i \to f {\tilde g}$ & - & - & 7 & 7  & 7\\  
 $\tilde f_i \to W^\pm \tilde f_j'$ & 7  & 7-8 & 8-9 & 8-9 & 8-9 \\ 
 $\tilde f_i \to H^\pm \tilde f_j'$ & 8  & 9-10 & 10-11 & 10-11 & 10-11\\
 $\tilde f_2 \to Z^0  \tilde f_1$   & 9  & - &  12 & 12 & -\\
 $\tilde f_2 \to A^0  \tilde f_1$   & 10 & - &  13 & 13 & -\\
 $\tilde f_2 \to h^0  \tilde f_1$   & 11 & - &  14 & 14 & -\\
 $\tilde f_2 \to H^0  \tilde f_1$   & 12 & - &  15 & 15 & -\\ \hline
 $\tilde l_1 \to l \, \tilde G$     & 13 & - & - & - & -\\ \hline
 $\tilde t_1 \to c {\tilde \chi}^0_{1,2}$ & - & - & - & - & 55-56 \\
 $\tilde t_1 \to W^+ \bar{b} {\tilde \chi}^0_1$ & - & - & - & - & 57 \\
 $\tilde t_1 \to \bar{b} e^+ {\tilde \nu}_e$ & - & - & - & - & 58 \\
 $\tilde t_1 \to \bar{b} \mu^+ {\tilde \nu}_\mu$ & - & - & - & - & 59 \\
 $\tilde t_1 \to \bar{b} \tau^+ {\tilde \nu}_\tau$ & - & - & - & - & 60 \\
 $\tilde t_1 \to \bar{b} \nu_e {\tilde e}^+_{1,2}$ & - & - & - & - & 61-62 \\
 $\tilde t_1 \to \bar{b} \nu_\mu {\tilde \mu}^+_{1,2}$ & -& -& -& -& 63-64 \\
 $\tilde t_1 \to \bar{b} \nu_\tau {\tilde \tau}^+_{1,2}$& -& -& -& -& 65-66 \\
\end{tabular}
\end{center}
\end{table}

\begin{table}
\caption{Correspondence between the second indices for chargino partial widths
(branching ratios) and the decay modes. }
\label{tab:charginodecays}
\begin{center}
\begin{tabular}{l|c}
mode & index of gP\_C ( BR\_C) \\ \hline
$\tilde \chi^+_i \to \tilde l_{m,k}^+ \, \nu_m$ & 1-6 \\ 
$\tilde \chi^+_i \to \tilde \nu_m \, l^+_m$ & 7-9 \\ 
$\tilde \chi^+_i \to \bar{\tilde d}_{m,k} \, u_m$ & 10-15 \\ 
$\tilde \chi^+_i \to \tilde u_{m,k} \, \bar{d}_m$ & 16-21 \\ 
$\tilde \chi^+_i \to \tilde \chi^0_j \, W^+$ & 22-25 \\
$\tilde \chi^+_i \to \tilde \chi^0_j \, H^+$ & 26-29 \\
$\tilde \chi^+_2 \to \tilde \chi^+_1 \, Z^0$ & 30 \\
$\tilde \chi^+_2 \to \tilde \chi^+_1 \, A^0$ & 31 \\
$\tilde \chi^+_2 \to \tilde \chi^+_1 \, h^0$ & 32 \\
$\tilde \chi^+_2 \to \tilde \chi^+_1 \, H^0$ & 33 \\ \hline
$\tilde \chi^+_i \to \tilde \chi^0_j \, u_m \, \bar{d}_m$ &  64-75 \\
$\tilde \chi^+_i \to \tilde \chi^0_j \, l^+_m \, \nu_m$ & 76-87 \\
$\tilde \chi^+_i \to \tilde g \, u_m \, \bar{d}_m$ & 88-90 \\
$\tilde \chi^+_2 \to \tilde \chi^+_1 \, u_m \, \bar{u}_m$ &  91-93 \\
$\tilde \chi^+_2 \to \tilde \chi^+_1 \, d_m \, \bar{d}_m$ &  94-96 \\
$\tilde \chi^+_2 \to \tilde \chi^+_1 \, l_m \, l^+_m$ &   97-99 \\
$\tilde \chi^+_2 \to \tilde \chi^+_1 \, \nu_m \, \bar{\nu}_m$ & 100-102 \\
\end{tabular}
\end{center}
\end{table}

\begin{table}
\caption{Correspondence between the second indices for neutralino partial 
widths
(branching ratios) and the decay modes. Note, that also charge conjugated
states are given.  }
\label{tab:neutralinodecays}
\begin{center}
\begin{tabular}{l|c}
mode & index of gP\_N ( BR\_N) \\ \hline
$\tilde \chi^0_i \to \tilde l_{m,k}^+ \, l_m$ & 1-12 \\ 
$\tilde \chi^0_i \to \tilde \nu_m \, \nu_m$ & 13-18 \\ 
$\tilde \chi^0_i \to \bar{\tilde u}_{m,k} \, u_m$ & 19-30 \\ 
$\tilde \chi^0_i \to \tilde d_{m,k} \, \bar{d}_m$ & 31-42 \\ 
$\tilde \chi^0_i \to \tilde \chi^\pm_j \, W^\mp$ & 43-46 \\
$\tilde \chi^0_i \to \tilde \chi^\pm_j \, H^\mp$ & 47-50 \\
$\tilde \chi^0_i \to \tilde \chi^0_j \, Z^0$ & 51-(24+i) \\
$\tilde \chi^0_i \to \tilde \chi^0_j \, A^0$ & (25+i)-(23+2 i) \\
$\tilde \chi^0_i \to \tilde \chi^0_j \, h^0$ & (26+i)-(22+3 i) \\
$\tilde \chi^0_i \to \tilde \chi^0_j \, H^0$ & (27+i)-(21+4 i) \\ \hline
$\tilde  \chi^0_i \to \gamma \, \tilde G$ & 63 \\
$\tilde  \chi^0_i \to  Z^0 \, \tilde G$ & 64 \\
$\tilde  \chi^0_i \to h^0 \, \tilde G$ & 65 \\ \hline
$\tilde  \chi^0_i \to \gamma \, \tilde  \chi^0_j$ & 65 + j \\
$\tilde \chi^0_i \to \tilde \chi^\pm_j \, q_m \, \bar{q}'_m$ & 69 - 80 \\
$\tilde \chi^0_i \to \tilde \chi^\pm_j \, l^\mp_m \, \nu_m$ & 81 - 92 \\
$\tilde \chi^0_i \to \tilde G  \, u_m \, \bar{u}_m$ & 93 - 95 \\
$\tilde \chi^0_i \to \tilde G  \, d_m \, \bar{d}_m$ & 96 - 98 \\
$\tilde \chi^0_{i>1} \to \tilde \chi^0_1 \, u_m \, \bar{u}_m$ & 99 - 101 \\
$\tilde \chi^0_{i>1} \to \tilde \chi^0_1 \, d_m \, \bar{d}_m$ & 102 - 104 \\
$\tilde \chi^0_{i>1} \to \tilde \chi^0_1 \, l^+_m \, l^-_m$ & 105 - 107 \\
$\tilde \chi^0_{i>1} \to \tilde \chi^0_1 \, \nu_m \, \bar{\nu}_m$
          & 108 - 110 \\
$\tilde \chi^0_{i>2} \to \tilde \chi^0_2 \, u_m \, \bar{u}_m$ & 111 - 113 \\
$\tilde \chi^0_{i>2} \to \tilde \chi^0_2 \, d_m \, \bar{d}_m$ & 114 - 116 \\
$\tilde \chi^0_{i>2} \to \tilde \chi^0_2 \, l^+_m \, l^-_m$ & 117 - 119 \\
$\tilde \chi^0_{i>2} \to \tilde \chi^0_2 \, \nu_m \, \bar{\nu}_m$
          & 120 - 122 \\
$\tilde \chi^0_4 \to \tilde \chi^0_3 \, u_m \, \bar{u}_m$ & 123 - 125 \\
$\tilde \chi^0_4 \to \tilde \chi^0_3 \, d_m \, \bar{d}_m$ & 126 - 128 \\
$\tilde \chi^0_4 \to \tilde \chi^0_3 \, l^+_m \, l^-_m$ & 129 - 131 \\
$\tilde \chi^0_4 \to \tilde \chi^0_3 \, \nu_m \, \bar{\nu}_m$
          & 132 - 134 \\
\end{tabular}
\end{center}
\end{table}

\begin{table}
\caption{Correspondence between the indices for gluino partial widths
(branching ratios) and the decay modes.  Note, that also charge conjugated
states are given. }
\label{tab:gluinodecays}
\begin{center}
\begin{tabular}{l|c}
mode & index of gP\_G ( BR\_G) \\ \hline
$\tilde g \to \bar{\tilde d}_{m,k} \, d_m$ & 1-12 \\ 
$\tilde g \to \tilde u_{m,k} \, \bar{u}_m$ & 13-24 \\ \hline 
$\tilde g \to \tilde t_1 \, \bar{c}$ & 25-26 \\ \hline
$\tilde g \to \gamma \, \tilde  \chi^0_j$ & 26 + j \\
$\tilde g \to \tilde \chi^0_1 \, u_m \, \bar{u}_m$ & 31-33 \\
$\tilde g \to \tilde \chi^0_1 \, d_m \, \bar{d}_m$ & 34-36 \\
$\tilde g \to \tilde \chi^0_2 \, u_m \, \bar{u}_m$ & 37-39 \\
$\tilde g \to \tilde \chi^0_2 \, d_m \, \bar{d}_m$ & 40-42 \\
$\tilde g \to \tilde \chi^0_3 \, u_m \, \bar{u}_m$ & 43-45 \\
$\tilde g \to \tilde \chi^0_3 \, d_m \, \bar{d}_m$ & 46-48 \\
$\tilde g \to \tilde \chi^0_4 \, u_m \, \bar{u}_m$ & 49-51 \\
$\tilde g \to \tilde \chi^0_4 \, d_m \, \bar{d}_m$ & 52-54 \\
$\tilde g \to \tilde \chi^\pm_i \, q_m \, \bar{q}_m'$ &  55-66 \\
$\tilde g \to \tilde t_i \, W^- \, \bar{b}$ &  67-68 \\
\end{tabular}
\end{center}
\end{table}
\begin{table}
\caption{Correspondence between the indices for the partial widths
(branching ratios) of the neutral Higgs bosons 
and the decay modes. The variables are gP\_S0 (BR\_S0) and  gP\_P0 (BR\_P0)
for the partial decay widths (branching ratios) of the CP-even Higgs
bosons ($h^0$, $H^0$) and CP-odd Higgs boson ($A^0$). In case of
gP\_S0 ( BR\_S0) the first (second) index denotes decay modes of $h^0$ ($H^0$).
Here $\phi$ stands for $h^0$, $H^0$ and $A^0$. The index m runs from 1 to 3.}
\label{tab:higgsdecays}
\begin{center}
\begin{tabular}{l|c|c|c} 
mode & $h^0$  & $H^0$  & $A^0$  \\ \hline
$\phi \to l^+_m \, l^-_m$ & 1-3 & 1-3 & 1-3 \\
$\phi \to d_m \, \bar{d}_m$ & 4-6 & 4-6 & 4-6 \\
$\phi \to u_m \, \bar{u}_m$ & 7-9 & 7-9 & 7-9 \\ \hline
$H^0 \to \tilde e^+_1 \, \tilde e^-_1$ & - & 10 & - \\
$\phi \to \tilde e^\mp_1 \, \tilde e^\pm_2$ & - & 11-12 & 11-12\\
$H^0 \to \tilde e^+_2 \, \tilde e^-_2$ & - & 13 & - \\
$H^0 \to \tilde \mu^+_1 \, \tilde \mu^-_1$ & - & 14 & - \\
$\phi \to \tilde \mu^\mp_1 \, \tilde \mu^\pm_2$ & - & 15-16 & 15-16\\
$H^0 \to \tilde \mu^+_2 \, \tilde \mu^-_2$ & - & 17 & - \\
$H^0 \to \tilde \tau^+_1 \, \tilde \tau^-_1$ & - & 18 & - \\
$\phi \to \tilde \tau^\mp_1 \, \tilde \tau^\pm_2$ & - & 19-20 & 19-20\\
$H^0 \to \tilde \tau^+_2 \, \tilde \tau^-_2$ & - & 21 & - \\
$H^0 \to \tilde \nu_{m} \, \overline{\tilde \nu}_m$ & - & 21 + m & - \\
$H^0 \to \tilde d^+_1 \, \tilde d^-_1$ & - & 25 & - \\
$\phi \to \tilde d^\mp_1 \, \tilde d^\pm_2$ & - & 26-27 & 23-24\\
$H^0 \to \tilde d^+_2 \, \tilde d^-_2$ & - & 28 & - \\
$H^0 \to \tilde s^+_1 \, \tilde s^-_1$ & - & 29 & - \\
$\phi \to \tilde s^\mp_1 \, \tilde s^\pm_2$ & - & 30-31 & 27-28\\
$H^0 \to \tilde s^+_2 \, \tilde s^-_2$ & - & 32 & - \\
$H^0 \to \tilde b^+_1 \, \tilde b^-_1$ & - & 33 & - \\
$\phi \to \tilde b^\mp_1 \, \tilde b^\pm_2$ & - & 34-35 & 31-32\\
$H^0 \to \tilde b^+_2 \, \tilde b^-_2$ & - & 36 & - \\
$H^0 \to \tilde u^+_1 \, \tilde u^-_1$ & - & 37 & - \\
$\phi \to \tilde u^\mp_1 \, \tilde u^\pm_2$ & - & 38-39 & 35-36\\
$H^0 \to \tilde u^+_2 \, \tilde u^-_2$ & - & 40 & - \\
$H^0 \to \tilde c^+_1 \, \tilde c^-_1$ & - & 41 & - \\
$\phi \to \tilde c^\mp_1 \, \tilde c^\pm_2$ & - & 42-43 & 39-40\\
$H^0 \to \tilde c^+_2 \, \tilde c^-_2$ & - & 44 & - \\
$H^0 \to \tilde t^+_1 \, \tilde t^-_1$ & - & 45 & - \\
$\phi \to \tilde t^\mp_1 \, \tilde t^\pm_2$ & - & 46-47 & 43-44\\
$H^0 \to \tilde t^+_2 \, \tilde t^-_2$ & - & 48 & - \\ \hline
$\phi \to \tilde \chi^0_r \, \tilde \chi^0_s$ $(r\le s)$
  & 49-58 & 49-58 &  46-55\\
$\phi \to \tilde \chi^+_1 \, \tilde \chi^-_1$& 59 & 59 &  56 \\
$\phi \to \tilde \chi^\pm_1 \, \tilde \chi^\mp_2$& 60-61 & 60-61 &  57-58 \\
$\phi \to \tilde \chi^+_2 \, \tilde \chi^-_2$& 62 & 62 &  59 \\ \hline
$H^0 \to Z^0 \, Z^0$ & - & 63 & - \\
$H^0 \to W^+ \, W^-$ & - & 64 & - \\
$H^0 \to h^0 \, h^0$ & - & 70 & - \\
$A^0 \to h^0 \, Z^0$ & - & - & 63 \\
\end{tabular}
\end{center}
\end{table}

\begin{table}
\caption{Correspondence between the indices for the partial widths
(branching ratios) of the charged Higgs 
and the decay modes. The variables are gP\_Spm ( BR\_Spm) and in case 
of partial decay widths (branching ratios). }
\label{tab:Hpdecays}
\begin{center}
\begin{tabular}{l|c}
mode & index \\  \hline
$H^+ \to l^+_m \, \nu_m$ & 1-3 \\
$H^+ \to u_m \, \bar{d}_m$ & 4-6 \\ \hline
$H^+ \to \tilde e^+_{i} \, \tilde \nu_e$ & 7-8 \\
$H^+ \to \tilde \mu^+_{i} \, \tilde \nu_\mu$ & 9-10 \\
$H^+ \to \tilde \tau^+_{i} \, \tilde \nu_\tau$ & 11-12 \\
$H^+ \to \tilde u_{i} \, {\overline{\tilde d}}_j$ & 12 + 2*(j-1) + i \\
$H^+ \to \tilde c_{i} \, {\overline{\tilde s}}_j$ & 16 + 2*(j-1) + i \\
$H^+ \to \tilde t_{i} \, {\overline{\tilde b}}_j$ & 20 + 2*(j-1) + i \\
$H^+ \to \tilde \chi^+_r  \, \tilde \chi^0_s$ & 24 + 4*(r-1) + s \\ \hline
$H^+ \to h^0 \, W^+$ & 34
\end{tabular}
\end{center}
\end{table}
Here the following variables are new:
\begin{itemize}
  \item \verb+epsI+ : gives the accuracy to which the phase space integrals
      in three body decays are calculated.
  \item \verb+deltaM+ : this variable affects the calculation of the  phase
       space integrals in three body decays. In case that 
       $m_i / (m - \sum_i m_i) < $ \verb+deltaM+ than $m_i$ is set to zero
       in the calculation of the phase space integrals.
       $m$ denotes here mass of the decaying particle and $m_i$ (i=1,2,3)
       are the masses of the decay products. 
  \item \verb+CalcTBD+ : if this variable is set \verb+.TRUE.+ then in all
      chargino-, neutralino- and gluino decays the three body modes will be
      calculated. This option has to be taken with care, because it can 
      slow down the program considerably.
  \item \verb+ratioWoM+ : this variable is used to decide whether two body
      decays or three body decay modes are calculated in case of charginos,
      neutralino and gluino. The program tries first two-body decay modes.
      In the case that the ratio of the width $\Gamma$ over the mass $m$ of the
      decaying particle is small: $\Gamma / m <$ \verb+ratioWoM+, then 
      three body decay modes are calculated.
  \item \verb+Fgmsb+ and \verb+m32+ : the $F$ parameter and the
    gravitino mass in the GMSB model. These parameters are calculated from
    the input and are set to huge numbers in all other models. They are
    needed for the calculation of the decay width(s) of the NLSP into a 
    gravitino. 
\end{itemize}

The next statements call the routine for the calculation of the cross sections
provided \verb+L_CS = .TRUE.+ and \verb+kont = 0+:
\begin{verbatim}
 If ((L_CS).and.(kont.eq.0)) then
  Call InitializeCrossSections(Ecms, Pm, Pp, ISR)
  Call CalculateCrossSections(Ecms, Pm, Pp, ISR                &
           & , mSup, RSup, mf_u, mSdown, RSdown, mf_d, mglu    &
           & , SigSup, SigSdown, mSlepton, RSlepton            &
           & , mSneut, RSneut, SigSle, SigSn, mC, U, V, mN, N  &
           & , SigC, SigChi0, mS0, RS0, vevSM, mP0, RP0, mSpm  &
           & , RSpm, SigS0, SigSP, SigHp )
 End If
\end{verbatim}
Here the following additional input is needed:
\begin{itemize}
 \item \verb+Ecms+ : the center of mass energy of the collider
 \item \verb+Pm+, \verb+Pp+ : degree of polarization of the incoming
    electron and positron, respectively
 \item \verb+ISR+ : logical variable, if \verb+.TRUE.+ then initial
  state radiation is taken into account using the formulas given in
  \cite{drees1}
\end{itemize}
These variables can be set in the file \verb+CrossSections.in+.
The cross sections are stored in the variables starting with \verb+Sig+
which are summarized in Table~\ref{tab:production}. Please note, that
in case of sfermions the structure of the variables is already put
such that the case of generation mixing can easily be implemented.
In the non--mixing case the cross sections are stored in the
$2\times 2$ diagonal blocks and they are sorted according to
the generations as in the case of the sfermion mixing matrices.
\begin{table}
\caption{Correspondence between the production cross sections
and the variables used in the program.}
\label{tab:production}
\begin{center}
\begin{tabular}{l|l}
process & Fortran name and type \\ \hline
$e^+ \, e^- \, \to \tilde u_i \, \tilde u_j$ & real(dp) :: SigSup(6,6) \\
$e^+ \, e^- \, \to \tilde d_i \, \tilde d_j$ & real(dp) :: SigSdown(6,6) \\
$e^+ \, e^- \, \to \tilde l_i \, \tilde l_j$ & real(dp) :: SigSle(6,6) \\
$e^+ \, e^- \, \to \tilde \nu_i \, \tilde \nu_j$ & real(dp) :: SigSn(6,6) \\
\hline
$e^+ \, e^- \, \to \tilde \chi^0_k \, \tilde \chi^0_n$
       & real(dp) :: SigChi0(4,4) \\
$e^+ \, e^- \, \to \tilde \chi^+_r \, \tilde \chi^-_s$
        & real(dp) :: SigN(4,4) \\ \hline
$e^+ \, e^- \, \to h^0 \, Z , \, \, H^0 Z$  & real(dp) :: SigS0(2) \\
$e^+ \, e^- \, \to h^0 \, A^0 , \, \, H^0 A^0$  & real(dp) :: SigSP(2) \\
$e^+ \, e^- \, \to  H^+ H^-$  & real(dp) :: SigHp
\end{tabular}
\end{center}
\end{table}

Finally the low energy constraints $b \to s \gamma$, $a_\mu$ and $\Delta \rho$
are calculated provided that calculation of the spectrum had been performed
successfully (\verb+kont.eq.0+):
\begin{verbatim}
 If (kont.eq.0) then
  Call CalculateLowEnergyConstraints(gauge, Y_l, Y_d, Y_u    &
    & , mSpm2, RSpm, mC, U, V, mN, N , mSup2, RSup, mSdown2  &
    & , RSdown, mSlepton2, RSlepton, mSneut2, RSneut         &
    & , BRbtosgamma, a_mu, Delta_Rho)
 Else
  BRbtosgamma = 0._dp
  a_mu = 0._dp
  Delta_Rho = 0._dp
 End If
\end{verbatim}
Here \verb+BRbtosgamma+, \verb+a_mu+, and \verb+Delta_Rho+ denote 
$10^4 \times \mathrm{BR}(b \to s \gamma)$, the SUSY contributions to
$a_\mu$ and the sfermion contributions to $\Delta \rho$, respectively.

Afterward the statement 
\begin{verbatim}
 Call WriteOutPut0(11, 1.e-6_dp, 1.e-3_dp)
\end{verbatim}
is used to write all information to the file connected with unit 11
(first entry). The second entry puts a lower bound on the branching ratios
 to be written. In the case above, branching ratios smaller
than $10^{-6}$ will not be given. The third entry gives the minimum value
for the cross section in $fb$ which will be written to the output file.
In the example given above cross sections smaller than $10^{-3}$ will not
be written to the output file.

The last statement closes all open files.
\begin{verbatim}
 call closing() ! closes the files
\end{verbatim}

\section{Conclusions}

We have described \verb+SPheno+, a program calculating the spectrum,
branching ratios and cross sections of supersymmetric particle in 
$e^+ e^-$ annihilation within the MSSM. The user can choose between the
following high scale models: minimal supergravity, minimal
supergravity including right handed neutrinos, gauge mediated
supersymmetry breaking, anomaly mediated supersymmetry breaking, and
string effective field theories based on OI and OII compactification.
 The calculation of the spectrum are done using two-loop renormalization group
equations and the complete one-loop formulas for the SUSY masses. 
In case of the neutral
Higgs bosons and the $\mu$ parameter leading two-loop effects are included.
The masses and mixing angles are used to calculate the most important
two body and three body decay modes.
They are also used for the  calculation of the SUSY  production cross sections
in $e^+ e^-$ annihilation. Here the effects of initial state radiation and
longitudinal beam polarization is included.
Finally the following low energy quantities are calculated:
$BR(b \to s \gamma)$, the supersymmetric contributions to the anomalous
magnetic moment of the muon $a_\mu$ and the sfermion contributions to
the $\rho$ parameter. Starting with version 2.2.0 \verb+SPheno+ 
allows for input and 
output according to the SUSY Les Houches accord \cite{Skands:2003cj}.

The program is set up in such a way that extensions can easily be 
included. The plans for upcoming versions are to include complex phases for
the supersymmetric parameters, to include generation mixing, to include
QCD and Yukawa corrections for various processes such as Sfermion and
Higgs production and decays. In addition beam strahlung for various
collider designs will be implemented.

\section*{Acknowledgments}
I would like to thank B.~Allanach, A.~Djouadi and S.~Kraml for discussions
on higher order corrections to  the Yukawa couplings. 
Special thanks go to: 
(i) P.~Slavich for discussions on 2-loop corrections of the neutral 
Higgs boson masses and the $\mu$ parameter as well as for providing a 
Fortran code performing these calculations.
(ii) M.~M\"uhlleitner for a detailed comparison of various decay branching 
ratios.
I would like to thank G.~Blair and P.~Zerwas for an enjoyable and fruitful
collaboration from which the work on this program originated.
This work is supported by the Erwin
Schr\"odinger fellowship No. J2095 of the `Fonds zur
F\"orderung der wissenschaftlichen Forschung' of Austria and
partly by the Swiss `Nationalfonds'.

\begin{appendix}

\section{Switches}
\label{app:switch}

In this appendix we describe the switches for influencing the behaviour
of the program.
Inside the main program one can set two scales:
\begin{enumerate}
 \item The electroweak scale $M_{EWSB}$, which is the scale where 
  the loop contributions to the masses and mixing matrices are calculated.
  The default is to calculate this scale from 
   $M_{EWSB}=\sqrt{m_{\tilde t_1} m_{\tilde t_2}}$. By calling
   \begin{verbatim}
  Call SetRGEScale(1.e3_dp**2)
    \end{verbatim}
   $M_{EWSB}$ will be set to the fixed value of $10^3$~GeV in this example.
   Note that the input is the scale squared.
   In the case one uses a zero or a negative number as input for
   \verb+SetRGEScale+ then the scale will be calculated from the
   stop masses.
 \item The high energy scale, where the boundary conditions of the model
   under study are set. Except for GMSB, where the scale is fixed by default,
   the high scale is calculated from the requirement $g_1 = g_2$. By calling
   \begin{verbatim}
 Call SetGUTScale(2.e16_dp)
   \end{verbatim}
    $M_{GUT}$ will be set to $2 \cdot 10^{16}$~GeV in this example. 
   In the case one uses a zero or a negative number as input for
   \verb+SetGUTScale+ then the scale will be calculated from the requirement
   $g_1 = g_2$ except for GMSB.
\end{enumerate}
In general the strong coupling $g_s$ will be different from $g_1$ and
$g_2$ in GUT theories if one works at the two loop level 
\cite{Weinberg:1980wa}. In case someone wants to enforce strict
universality at the high scale, this can be done by using the following
statement:
\begin{verbatim}
 test = SetStrictUnification(.TRUE.)
\end{verbatim}
where \verb+test+ is a logical variable to which the old value of
SetStrictUnification is assigned. The same functions can also be used
to set this option \verb+.FALSE.+.

Starting with version 2.2.0 it also possible to set these values using
the SUSY Les Houches accord as described in Appendix~\ref{app:SLHA}.

\section{Input files}
In this section the input files are described. Among these files only
the file \verb+HighScale.in+ has to be provided by the user. The other
files \verb+Control.in+, \verb+CrossSections.in+
 and \verb+StandardModel.in+ can be used to change
the default values which are given below in Appendices~\ref{app:Control},
\ref{app:cross} and \ref{app:SM}. An alternative way to provide \verb+SPheno+
with input is by the SUSY Les Houches accord as described in 
Appendix~\ref{app:SLHA}.

\subsection{Control.in}
\label{app:Control}

This file contains three entries as shown below:
\begin{verbatim}
0             ! ErrorLevel
.True.        ! Calculation of branching ratios
.True.        ! Calculation of cross sections
\end{verbatim}
The values given above are the default values inside the program and are
used if the the file \verb+Control.in+ is not present. Here 
\verb+ErrorLevel+ is an integer in the range [-2,2] where the numbers
correspond to the following behaviour of the program:
\begin{enumerate}
 \item[-2] do not print severe warnings
 \item[-1]  do not print warnings
 \item[0]  print every warning
 \item[1] abort in case of a severe warning
 \item[2]  abort even in case of a warning
\end{enumerate}
A warning is called severe if either a result is unphysical or if
a numerical procedure gives an unreliable result.

\subsection{CrossSections.in}
\label{app:cross}

This file contains four entries as shown below:
\begin{verbatim}
500.          ! c.m.s. energy in GeV
0.            ! degree of longitudinal polarization of electrons
0.            ! degree of longitudinal polarization of positrons
.True.        ! calculation of initial state radiation if .TRUE. 
\end{verbatim}
The values given above are the default values inside the program and are
used if the the file \verb+CrossSections.in+ is not present. 

\subsection{HighScale.in}
\label{app:model}
In this section we describe the input file for the high scale boundary
conditions. The package contains several files starting with 
\verb+Highscale.in+ and ending in the models described below. One has
to rename the model file to the name \verb+Highscale.in+ to use it as
input for \verb+SPheno+. Note that in all examples below the value
of $A_0$ given below will be multiplied by the Yukawas coupling at the
high scale and that this product enters the RGEs. For most of the
examples below we have used the so-called SPS points defined in 
\cite{Allanach:2002nj}. In Appendix~\ref{app:output} we display the output
for the point SPS1a.

\subsubsection{mSUGRA}
\label{app:mSUGRA}

The minimal SUGRA version is defined by four parameters and the sign of
the $\mu$ parameter. The parameters are the gaugino mass parameter
$M_{1/2}$, the scalar mass parameter $M_0$, the trilinear parameter
$A_0$ as well as $\tan\beta$. The file reads for example
\begin{verbatim}
mSugra
 250.                         ! M_1/2
 100.                         ! M_0
-100.                         ! A_0
  10.                         ! tan(beta)
   1.                         ! sign of mu
.TRUE.                        ! if 2-loop RGEs should be used
\end{verbatim}

\subsubsection{mSUGRA including right handed neutrinos}
\label{app:mSUGRAnuR}

In this case one needs four more input values compared to the case of
mSUGRA described above: a common right handed neutrino mass $m_{\nu_R}$
and the light neutrino masses $m_{\nu_i}$ ($i=1,2,3$).
The file reads for example
\begin{verbatim}
mSugra
 250.                         ! M_1/2
 100.                         ! M_0
-100.                         ! A_0
  10.                         ! tan(beta)
   1.                         ! sign of mu
1.e14                         ! m_nu_R
1.e-14 3.e-12 0.06e-9         ! m_nu_i 
.TRUE.                        ! if 2-loop RGEs should be used
\end{verbatim}

\subsubsection{GMSB}

The GMSB scenario is characterized by two mass parameters $M_M$ and
$\Lambda$; the multiplicity $N_5$ and $N_{10}$ of  messengers in the
$5+\overline{5}$ and $10+\overline{10}$ vector-like multiplets, respectively;
$\tan\beta$ and the sign of $\mu$  as described in \sect{sect:gmsb}.
In addition one can set a common value for the $A$ parameters at the scale
$M_M$. Note, however, that in the minimal model this value is practically
zero. The file reads for example
\begin{verbatim}
GMSB
100000.                        ! Lambda
200000.                        ! M_M
1                              ! N_5
0                              ! N_10
0.                             ! A0
15.                            ! tan(beta)
1.                             ! sign of mu
.TRUE.                         ! if 2-loop RGEs should be used
\end{verbatim}

\subsubsection{AMSB}

The implemented AMSB scenario is characterized by the gravitino mass
$m_{3/2}$, a common scalar mass $M_0$, $\tan\beta$ and the sign of $\mu$.
 The file reads for example
\begin{verbatim}
AMSB            ! model
60000.          ! M_3/2
450.            ! M_0
10.             ! tan(beta)
1.              ! sign of mu
.TRUE.          ! if 2-loop RGEs should be used
\end{verbatim}

\subsubsection{String I}

This scenario is characterized by the gravitino mass $m_{3/2}$, the
common vev $<t>$ of the moduli fields, the string coupling squared
$g^2_s$, the sine squared of the mixing angle between the dilaton fields
and moduli fields $\sin^2 \theta$, the parameter $\delta_{GS}$
 of the Green--Schwarz
counter-term, the modular weights $n_i$ characterizing
the couplings between moduli fields and matter fields, which are assumed
to be generation independent in the current implementation. Moreover,
one needs to specify $\tan\beta$ and the sign of $\mu$. The file reads for
example as
\begin{verbatim}
String_OI
180.                      ! M_3/2
14.0                      ! <t>
0.5                       ! g_s^2
0.9                       ! sin^2(theta)
 0.                       ! delta_GS
-1 -3                     ! n_E n_L
 1 -2 0                   ! n_D n_U n_Q
-1 -1                     ! n_H1 n_H2
10.                       ! tan(beta)
-1.                       ! phase(mu)
.TRUE.                    ! if 2-loop RGEs should be used
\end{verbatim}

\subsubsection{String II}

This scenario is characterized by the gravitino mass $m_{3/2}$, the
common vev of the moduli fields $<t>$, the string coupling squared
$g^2_s$, the sine squared of the mixing angle between the dilaton fields
and moduli fields $\sin^2 \theta$, the parameter of the Green--Schwarz
counter-term $\delta_{GS}$. Moreover,
one needs to specify $\tan\beta$ and the sign of $\mu$. There are two
different scenarios implemented denoted as \verb+String_OIIa+ and
 \verb+String_OIIb+ corresponding to boundary conditions $(A)$ and
$(B)$ of \cite{Binetruy:2001md}, respectively.
The files read for example as
\begin{verbatim}
String_OIIa
300.                      ! M_3/2
14.6                      ! <t>
0.5                       ! g_s^2
0.9                       ! sin^2(theta)
0.                        ! delta_GS
5.                        ! tan(beta)
1.                        ! phase(mu)
.TRUE.                    ! if 2-loop RGEs should be used
\end{verbatim}
and
\begin{verbatim}
String_OIIb
300.                      ! M_3/2
14.6                      ! <t>
0.5                       ! g_s^2
0.9                       ! sin^2(theta)
0.                        ! delta_GS
5.                        ! tan(beta)
1.                        ! phase(mu)
.TRUE.                    ! if 2-loop RGEs should be used
\end{verbatim}

\subsubsection{SUGRA}

This input files serves as interface for more general models with
gauge couplings unification. Here the user can specify non-universal
gaugino masses at the high scale $M_{1/2}[U(1)]$, $M_{1/2}[SU(2)]$,
$M_{1/2}[SU(3)]$, 15 different values of the sfermion mass parameters for 
every type of sfermions: $M^0_{\tilde E,ii}$, $ M^0_{\tilde L,ii}$, 
$ M^0_{\tilde D,ii}$,
$ M^0_{\tilde U,ii}$, $ M^0_{\tilde Q,ii}$; two Higgs mass parameters
$M^0_{H_1}$ and $M^0_{H_2}$; nine different $A$ parameters $A_{0,e,ii}$,
 $A_{0,d,ii}$ and  $A_{0,u,ii}$. Here $ii$ denotes that only the
diagonal entries can be set.

\begin{verbatim}
Sugra
480. 300. 300.       ! M_1/2_i
150. 150. 150.       ! M0_E_ii
150. 150. 150.       ! M0_L_ii
150. 150. 150.       ! M0_D_ii
150. 150. 150.       ! M0_Q_ii
150. 150. 150.       ! M0_U_ii
150. 150.            ! M0_H_i
0. 0. 0.             ! A0_u_ii
0. 0. 0.             ! A0_d_ii
0. 0. 0.             ! A0_e_ii
10.                  ! tan(beta)
1.                   ! phase(mu)
.TRUE.               ! if 2-loop RGEs should be used
\end{verbatim}

\subsubsection{MSSM}

This input file serves for the cases that one wants to start with 
low energy parameters for the calculation of masses, decays and/or production
rates. The input consists of the electroweak gaugino mass parameters $M_1$ and
$M_2$, the gluino mass $m_{\tilde g}$,  15 sfermion mass parameters
$M_{\tilde E,ii}$, $M_{\tilde L,ii}$, $M_{\tilde D,ii}$, $M_{\tilde U,ii}$, 
$M_{\tilde Q,ii}$, 9 trilinear parameters $A_{e,ii}$, $A_{d,ii}$, $A_{u,ii}$;
$\tan\beta$. Moreover, one has to specify the renormalisation scale $Q$ where
all the parameters are given as well as $\mu$ and the mass of the pseudoscalar
Higgs $m_A$.
\begin{verbatim}
MSSM
99.13 192.74 580.51                         ! M_1 M_2 m_gluino
136.23 136.23 133.55                        ! M_E_i
196.64 196.64 195.75                        ! M_L_i
519.53 519.53 516.86                        ! M_D_i
539.86 539.86 495.91                        ! M_Q_i
521.66 521.66 424.83                        ! M_U_i
0. 0. -510.01                               ! A_u
0. 0. -772.66                               ! A_d
0. 0. -254.20                               ! A_e
10. 454.65                                  ! tan(beta) Q
352.39  393.63                              ! mu m_A
\end{verbatim}
In the case that the keyword \verb+MSSMtree+ is used instead of \verb+MSSM+
the masses are calculated using tree-level formulas instead of loop corrected
masses. The renormalisation scale Q is absent in this case, e.g.~ the 
two last line read as
\begin{verbatim}
10.                                         ! tan(beta)
352.39  393.63                              ! mu m_A
\end{verbatim}

\subsection{StandardModel.in}
\label{app:SM}

This file contains the values of the Standard Model parameters and must
include all lines given below. Otherwise the default values given
in the listing below are used: 
\begin{verbatim}
91.1876                   ! m_Z
2.4952                    ! Gamma_Z
0.0336 0.0336 0.037       ! Br(Z -> l l)
0.2                       ! Br(Z -> invisible)
2.118                     ! Gamma_W
0.1  0.1  0.1             ! Br(W -> l nu)
0.51099890e-3             ! m_e
0.105658357               ! m_mu
1.7770                    ! m_tau
2.00             ! scale Q where the masses of the light quarks u,d,s,c are given
0.003            ! m_u(Q)
1.2              ! m_c(Q)
175.0            ! m_t, pole mass
0.007            ! m_d(Q)
0.12             ! m_s(Q)
4.25             ! m_b(m_b)
137.0359998      ! 1./ alpha
0.1172           ! alpha_s(m_Z)
1.16639e-5       !  G_F, Fermi constant
0.224            ! s12 of CKM particle data book 1998, 90%, 0.217-0.222
0.0413           ! s23 of CKM 0.036-0.042
0.00363           ! s13 of CKM 0.0018 - 0.0044
0.               ! phase of CKM   0.-2 Pi
2.19709e-6       ! life time of muon
3.4e-13          ! life time of tau
\end{verbatim}
All masses are given in GeV and
$Br$ denotes ``branching ratio'' in the list above.

\section{Implementation of SUSY Les Houches Accord}
\label{app:SLHA}

Starting with version 2.2.0 \verb+SPheno+ allows for input and output according
to the SUSY Les Houches accord \cite{Skands:2003cj}. The name of the
input file is \verb+LesHouches.in+ and the output will be written to the
file \verb+SPheno.spc+.
In the following we summarize
unsupported features as well extensions of this standard. The unsupported
features are:
\begin{itemize}
 \item $\alpha^1_{em}(m_Z)^{\overline{MS}}$ of the block \verb+SMINPUTS+
   is ignored as input. Instead
   the value $\alpha_{em}(0)$ can be changed in the block 
   \verb+SPhenoInput+ described below. The corresponding value for the
   output will be given.
 \item In block \verb+EXTPAR+ the entries 51--53 are ignored as the
   corresponding formulas are not (yet) implemented in \verb+SPheno+.
 \item Currently there is no information concerning warnings and errors
     in block \verb+SPINFO+. This will be changed within the one of the
     next versions. Please check the file \verb+Messages.out+ for this
     information.
\end{itemize}
The current implementation requires that the block \verb+MODSEL+ is read
in before the block \verb+EXTPAR+ is read in. In the case that an unknown
entry appears, a warning message is printed and SPheno tries to proceed.
SPheno stops execution in the case that the model input is not complete.

For the \verb+SPheno+ specific input the block \verb+SPhenoInput+ has to be 
used.
In this block switches can be set, SM input beside the one of the block 
\verb+SMINPUTS+ can be set. Moreover, the information for the cross section
calculation can be given here. Within this block the following 
flags and parameters can be set, with general structure \verb+id value+: \\
\begin{tabular}{rcl}
\tt 1 &:& setting the error level as described Appendix~\ref{app:Control} \\
\tt 11 &:& if value=1 (0) then (no) branching ratios are calculated \\
\tt 12 &:& only branching ratios larger than value are written out \\
\tt 21 &:& if value=1 (0) then (no) cross sections are calculated \\
\tt 22 &:& cms energy for $e^+ e^-$ annihilation \\
\tt 23 &:& value gives degree of polarisation for $e^-$ beam \\
\tt 24 &:& value gives degree of polarisation for $e^+$ beam \\
\tt 25 &:& only cross sections larger than value (in fb) are written out \\
\tt 31 &:& a fixed value for the GUT scale is used if value is larger than 0 \\
\tt 32 &:& if value=0 then $g_3(m_{GUT})$ can be different from 
           $g_1(m_{GUT})=g_2(m_{GUT})$; \\
       && if value=1 then strict unification
           $g_1(m_{GUT})=g_2(m_{GUT})=g_3(m_{GUT})$ is enforced \\
\tt 33 &:& a fixed value for the renormalization scale $Q_{EWSB}$
           is used if value is larger than 0 \\
\tt 41 &:& sets value of Z-boson width  \\
\tt 42 &:& sets value of W-boson width  \\
\tt 51 &:& sets value of electron mass  \\
\tt 52 &:& sets value of muon mass  \\
\tt 61 &:& sets scale where the running masses for light quarks ($u,d,s,c$)
           are defined \\
\tt 62 &:& sets value of u-quark mass \\
\tt 63 &:& sets value of c-quark mass \\
\tt 64 &:& sets value of d-quark mass \\
\tt 65 &:& sets value of s-quark mass \\
\end{tabular}

\noindent
Here is an example for this block with the default values of \verb+SPheno+
\begin{verbatim}
Block SPhenoInput       # SPheno specific input
 1  -1                  # error level
11   1                  # calculate branching ratios
12   1.00000000E-04     # write only branching ratios larger than this value
21   1                  # calculate cross section 
22   5.00000000E+02     # cms energy in GeV
23   0.00000000E+00     # polarisation of incoming e- beam
24   0.00000000E+00     # polarisation of incoming e+ beam
25   1.00000000E-04     # write only cross sections larger than this value [fb]
31  -1.00000000E+00     # m_GUT, if < 0 than it determined via g_1=g_2
32   0                  # require strict unification g_1=g_2=g_3 if '1' is set
33  -1.00000000E+00     # Q_EWSB, if < 0 than  Q_EWSB=sqrt(m_~t1 m_~t2)
41   2.49520000E+00     # width of the Z-boson
42   2.11800000E+00     # width of the W-boson
51   5.10998900E-04     # electron mass
52   1.05658357E-01     # muon mass
61   2.00000000E+00     # scale where quark masses of first 2 gen. are defined
62   3.00000000E-03     # m_u(Q)
63   1.20000000E+00     # m_c(Q)
64   7.00000000E-03     # m_d(Q)
65   1.20000000E-01     # m_s(Q)
\end{verbatim}
 
For the output the extensions below have been defined.
The information concerning the cross section is written out
using a \verb+SPheno+ specific block called
\verb+SPhenoCrossSections+. The first line of this block gives the
information on the cms energy, the polarization of the incoming beams
as well if ISR is included or not, for example for $\sqrt{s}=500$~GeV and
unpolarized beams:
\begin{verbatim}
Block SPhenoCrossSections  # cross sections
XS 11 -11   500.0  0.00  0.00  1  # e+ e- XS, Pe-, Pe+,  including ISR
\end{verbatim}
The FORTRAN format is in this case:
\begin{verbatim}
Format("XS 11 -11 ",F7.1," ",F5.2," ",F5.2," ",A)
\end{verbatim}
The cross sections (in fb) themself are written as
\begin{verbatim}
#      Sigma [fb]       NDA        ID1     ID2
     2.83574498E+02    2     2000011  -2000011    # ~e_R-      ~e_R+
\end{verbatim}
Here the first entry gives the cross section in fb, the second entry specifies
the number of produced particles, the subsequent two integers give the
PDG code of the particles.
We have used the FORTRAN format
\begin{verbatim}
Format(3x,1P,e16.8,0p,3x,I2,3x,2(i9,1x),2x," # ",A)
\end{verbatim}
As an example we give  the cross sections for the SPS1a scenario
at an 500 GeV $e^+ e^-$ linear collider with unpolarized beams:
\begin{verbatim}
Block SPhenoCrossSections  # cross sections
XS 11 -11   500.0  0.00  0.00  1  # e+ e- XS, Pe-, Pe+,  including ISR
#      Sigma [fb]       NDA        ID1     ID2
     2.83574498E+02    2     2000011  -2000011    # ~e_R-      ~e_R+
     7.79728001E+01    2     2000011  -1000011    # ~e_R-      ~e_L+
     4.57495061E+01    2     1000011  -1000011    # ~e_L-      ~e_L+
     5.47916441E+01    2     2000013  -2000013    # ~mu_R-     ~mu_R+
     6.00045490E-03    2     2000013  -1000013    # ~mu_R-     ~mu_L+
     1.90114309E+01    2     1000013  -1000013    # ~mu_L-     ~mu_L+
     5.96228076E+01    2     1000015  -1000015    # ~tau_1-    ~tau_1+
     1.26426385E+00    2     1000015  -2000015    # ~tau_1-    ~tau_2+
     1.59684572E+01    2     2000015  -2000015    # ~tau_2-    ~tau_2+
     4.52889205E+02    2     1000012  -1000012    # ~nu_eL     ~nu_eL*
     1.36168303E+01    2     1000014  -1000014    # ~nu_muL    ~nu_muL*
     1.39168830E+01    2     1000016  -1000016    # ~nu_tauL   ~nu_tauL*
     2.75869582E+02    2     1000022   1000022    # chi_10 chi_10
     6.56937491E+01    2     1000022   1000023    # chi_10 chi_20
     7.10141133E+00    2     1000022   1000025    # chi_10 chi_30
     8.27993814E-01    2     1000022   1000035    # chi_10 chi_40
     6.90281358E+01    2     1000023   1000023    # chi_20 chi_20
     1.60903760E+02    2     1000024  -1000024    # chi_1- chi_1+
     2.47077869E+01    2          25        23    # h0 Z
\end{verbatim}

The information concerning the value of low energy observables 
($BR(b\to s \gamma)$, SUSY contribution to $(g-2)_\mu$ and $\Delta(\rho)$)
is written out
using a \verb+SPheno+ specific block called
\verb+SPhenoLowEnergy+. We use the following identifier:\\
\begin{tabular}{rcl}
\tt 1 &:& BR$(b\to s \gamma)$ \\
\tt 2 &:& SUSY contribtutions to  $(g-2)_\mu$ \\
\tt 3 &:& SUSY contribtutions to $\Delta(\rho)$ \\
\end{tabular}\\
As an example we give here the output for the SPS1a scenario:
\begin{verbatim}
Block SPhenoLowEnergy  # low energy observables
    1    4.55809155E+00   # BR(b -> s gamma)
    2    5.42193822E-09   # (g-2)_muon
    3    1.97608480E-04   # Delta(rho)
\end{verbatim}

\section{Sample output}
\label{app:output}
Here we give the content of the file \verb+SPheno.out+ provided one
uses the content of \verb+HighScale.in+ for the mSUGRA scenario
described in Appendix~\ref{app:mSUGRA}
and the default values of the files \verb+Control.in+, \verb+CrossSections.in+
and \verb+StandardModel.in+.

\begin{verbatim}
SPheno output file
Version 2.2.0 ,  created: 11.03.2004,  21:06
  
 mSugra input at the GUT scale   2.3626654756304632E+16
 M_1/2     :    2.5000000000000000E+02
 M_0       :    1.0000000000000000E+02
 A_0       :   -1.0000000000000000E+02
 tan(beta) at m_Z :   10.0000000000000000
 phase(mu) :    1.0000000000000000
  
 Parameters at the scale    4.8391963443672887E+02
  
        g'             g             g_3
  3.61193779E-01  6.46479958E-01  1.09493111E+00
  
       Y_e            Y_mu          Y_tau
  2.88471294E-05  5.96467232E-03  1.00322460E-01
  
       Y_u            Y_c            Y_t
  8.84695152E-06  3.53878053E-03  8.92107505E-01
  
       Y_d            Y_s            Y_b
  1.91375025E-04  3.28071453E-03  1.37357260E-01
  
 Gaugino mass parameters
   1.0165957170339030E+02   1.9176076501787236E+02   5.8493277901867339E+02
  
 mu, B
   3.5735299500991079E+02   1.6699996261573800E+04
  
 Slepton mass parameters
 A_l
  -2.5325613989602988E+02  -2.5325004783357019E+02  -2.5153292012959056E+02
 M2_E
   1.8440768268311054E+04   1.8438451866710398E+04   1.7787517468332015E+04
 M2_L
   3.8191007905480838E+04   3.8189867711691244E+04   3.7869523746231192E+04
  
 Squark mass parameters
 A_d
  -8.5467574322023552E+02  -8.5467228770510383E+02  -7.9058921072895021E+02
 A_u
  -6.7949078243535337E+02  -6.7948707229660602E+02  -4.9657599289477690E+02
 M2_D
   2.7447722422748851E+05   2.7447520834575844E+05   2.7114817843756522E+05
 M2_U
   2.7672441991504660E+05   2.7672246826518845E+05   1.7613291366656474E+05
 M2_Q
   2.9674725559000188E+05   2.9674529488349473E+05   2.4602686847035179E+05
  
 Higgs mass parameters
   3.2530345774801379E+04  -1.2784263452390164E+05
  
 Masses and mixing matrices
 Gluino :    6.0402198802780606E+02   1.0000000000000000
  
 Charginos
   1.7973283139106542E+02   3.8228031192319548E+02
   U
         -0.91793         0.39673
          0.39673         0.91793
   V
         -0.97122         0.23819
          0.23819         0.97122
  
 Neutralinos
  97.1655612758851674   1.8073254909635224E+02   3.6466658683987458E+02   3.8192617586445749E+02
   N
    1   1  (        -0.98576,         0.00000)
    1   2  (         0.05614,         0.00000)
    1   3  (        -0.14885,         0.00000)
    1   4  (         0.05449,         0.00000)
    2   1  (        -0.10391,         0.00000)
    2   2  (        -0.94275,         0.00000)
    2   3  (         0.27483,         0.00000)
    2   4  (        -0.15776,         0.00000)
    3   1  (         0.00000,         0.06048)
    3   2  (        -0.00000,        -0.09026)
    3   3  (        -0.00000,        -0.69484)
    3   4  (        -0.00000,        -0.71091)
    4   1  (         0.11757,         0.00000)
    4   2  (        -0.31610,         0.00000)
    4   3  (        -0.64770,         0.00000)
    4   4  (         0.68319,         0.00000)
  
 e-sneutrino mass   :    1.9127930364369408E+02
 mu-sneutrino mass  :    1.9127614461238977E+02
 tau-sneutrino mass :    1.9038540593550744E+02
  
 selectron masses
   1.4393792438692086E+02   2.0712432781915908E+02
  R_e
          0.00009         1.00000
         -1.00000         0.00009
  
 smuon masses
   1.4390451752393614E+02   2.0713856716885780E+02
  R_mu
          0.01790         0.99984
         -0.99984         0.01790
  
 stau masses
   1.3484343437971901E+02   2.1068049221898704E+02
  R_tau
          0.26425         0.96445
         -0.96445         0.26425
  
 u-squark masses
   5.4756917879839784E+02   5.6524131281347229E+02
  R_u
          0.00006         1.00000
         -1.00000         0.00006
  
 c-squark masses
   5.4755738858800510E+02   5.6524965433179352E+02
  R_c
          0.02389         0.99971
         -0.99971         0.02389
  
 t-squark masses
   3.9945389533787068E+02   5.8628028438847809E+02
  R_t
          0.55340         0.83291
         -0.83291         0.55340
  
 d-squark masses
   5.4730506514994352E+02   5.7064687850962798E+02
  R_d
          0.00058         1.00000
         -1.00000         0.00058
  
 s-squark masses
   5.4730068877040878E+02   5.7064721260745614E+02
  R_s
          0.01000         0.99995
         -0.99995         0.01000
  
 b-squark masses
   5.1512514815767906E+02   5.4710587491028832E+02
  R_b
          0.94697         0.32132
         -0.32132         0.94697
  
 m_A0, m_H+
   3.9942973783298265E+02   4.0772744504014719E+02
  
 m_h0, m_H0
   1.1080888761166447E+02   3.9980703384665037E+02
  R_S0
          0.11371         0.99351
         -0.99351         0.11371
  
  
 Low energy constraints 
  10^4 Br(b -> s gamma) :   0.4558514E+01
  Delta(a_mu)           :   0.5415330E-08
  Delta(rho)            :   0.1959489E-03
  
  
  Anti particles are marked with a * in case of
  (s)neutrinos and (s)quarks in the decay section.
                     Decay widths (GeV) and branching ratios
  
  Selectron_1
   Neutralino_1 e               0.21510088  100.00000000
  Total width :                 0.21510088
  
  
  Selectron_2
   Neutralino_1 e               0.12813720   48.21320960
   Neutralino_2 e               0.04910231   18.47535119
   Chargino_1 neutrino          0.08853247   33.31143921
  Total width :                 0.26577199
  
  
  Smuon_1
   Neutralino_1 mu              0.21484110  100.00000000
  Total width :                 0.21484110
  
  
  Smuon_2
   Neutralino_1 mu              0.12829264   48.23209605
   Neutralino_2 mu              0.04912578   18.46902142
   Chargino_1 neutrino          0.08857176   33.29888253
  Total width :                 0.26599018
  
  
  Stau_1
   Neutralino_1 tau             0.15075247  100.00000000
  Total width :                 0.15075247
  
  
  Stau_2
   Neutralino_1 tau             0.16323638   51.63787855
   Neutralino_2 tau             0.05485389   17.35237248
   Chargino_1 neutrino          0.09802725   31.00974897
  Total width :                 0.31611751
  
  
  e-Sneutrino
   Neutralino_1 neutrino        0.16123722   85.32866208
   Neutralino_2 neutrino        0.00715724    3.78769424
   Chargino_1 e                 0.02056576   10.88364368
  Total width :                 0.18896022
  
  
  mu-Sneutrino
   Neutralino_1 neutrino        0.16123086   85.33568036
   Neutralino_2 neutrino        0.00715318    3.78600969
   Chargino_1 mu                0.02055318   10.87830995
  Total width :                 0.18893721
  
  
  tau-Sneutrino
   Neutralino_1 neutrino        0.15943476   87.31209427
   Neutralino_2 neutrino        0.00605108    3.31378333
   Chargino_1 tau               0.01711746    9.37412241
  Total width :                 0.18260329
  
  
  Sdown_1
   Neutralino_1 d-quark         0.28770802   98.55660629
   Neutralino_2 d-quark         0.00270653    0.92714381
   Neutralino_3 d-quark         0.00035708    0.12232218
   Neutralino_4 d-quark         0.00114841    0.39339795
   Chargino_1 u-quark           0.00000154    0.00052910
  Total width :                 0.29192159
  
  
  Sdown_2
   Neutralino_1 d-quark         0.12854555    2.41263613
   Neutralino_2 d-quark         1.63736438   30.73124285
   Neutralino_3 d-quark         0.00855909    0.16064329
   Neutralino_4 d-quark         0.08259529    1.55020837
   Chargino_1 u-quark           3.24403336   60.88637215
   Chargino_2 u-quark           0.22691457    4.25889723
  Total width :                 5.32801224
  
  
  S-strange_1
   Neutralino_1 s-quark         0.28771851   98.32632702
   Neutralino_2 s-quark         0.00292980    1.00124494
   Neutralino_3 s-quark         0.00036645    0.12523375
   Neutralino_4 s-quark         0.00114683    0.39192166
   Chargino_1 c-quark           0.00045378    0.15507723
   Chargino_2 c-quark           0.00000057    0.00019540
  Total width :                 0.29261594
  
  
  S-strange_2
   Neutralino_1 s-quark         0.12853671    2.41273081
   Neutralino_2 s-quark         1.63714210   30.73039024
   Neutralino_3 s-quark         0.00858872    0.16121673
   Neutralino_4 s-quark         0.08262979    1.55102333
   Chargino_1 c-quark           3.24352858   60.88347439
   Chargino_2 c-quark           0.22701084    4.26116451
  Total width :                 5.32743673
  
  
  Sbottom_1
   Neutralino_1 b-quark         0.16674913    4.31705130
   Neutralino_2 b-quark         1.34234704   34.75269181
   Neutralino_3 b-quark         0.01963712    0.50839523
   Neutralino_4 b-quark         0.04237541    1.09707817
   Chargino_1 t-quark           1.72934949   44.77199120
   Stop_1 W-                    0.56211179   14.55279229
  Total width :                 3.86256998
  
  
  Sbottom_2
   Neutralino_1 b-quark         0.23969032   31.70912297
   Neutralino_2 b-quark         0.09310073   12.31648674
   Neutralino_3 b-quark         0.04221537    5.58475713
   Neutralino_4 b-quark         0.05849015    7.73778189
   Chargino_1 t-quark           0.12448615   16.46852800
   Stop_1 W-                    0.19792061   26.18332328
  Total width :                 0.75590333
  
  
  Sup_1
   Neutralino_1 u-quark         1.15145952   98.55637070
   Neutralino_2 u-quark         0.01083087    0.92704170
   Neutralino_3 u-quark         0.00143110    0.12249171
   Neutralino_4 u-quark         0.00460431    0.39409473
  Total width :                 1.16832581
  
  
  Sup_2
   Neutralino_1 u-quark         0.03594243    0.65176659
   Neutralino_2 u-quark         1.75313101   31.79062555
   Neutralino_3 u-quark         0.00499759    0.09062447
   Neutralino_4 u-quark         0.06006824    1.08925513
   Chargino_1 d-quark           3.58197347   64.95417438
   Chargino_2 d-quark           0.07850353    1.42355387
  Total width :                 5.51461627
  
  
  S-charm_1
   Neutralino_1 c-quark         1.15081537   98.28078206
   Neutralino_2 c-quark         0.01188610    1.01508491
   Neutralino_3 c-quark         0.00147943    0.12634444
   Neutralino_4 c-quark         0.00458850    0.39186267
   Chargino_1 s-quark           0.00217691    0.18591022
  Total width :                 1.17094649
  
  
  S-charm_2
   Neutralino_1 c-quark         0.03656941    0.66343674
   Neutralino_2 c-quark         1.75204300   31.78529687
   Neutralino_3 c-quark         0.00499658    0.09064717
   Neutralino_4 c-quark         0.06014129    1.09107406
   Chargino_1 s-quark           3.57978849   64.94397663
   Chargino_2 s-quark           0.07857902    1.42556852
  Total width :                 5.51211779
  
  
  Stop_1
   Neutralino_1 t-quark         0.39740101   19.20680823
   Neutralino_2 t-quark         0.24141364   11.66777483
   Chargino_1 b-quark           1.39258046   67.30487641
   Chargino_2 b-quark           0.02117232    1.02328034
   c-quark neutralino_1         0.00039523    0.01910206
   c-quark neutralino_2         0.01609207    0.77774681
   W b-quark neutralino_1       0.00000851    0.00041131
  Total width :                 2.06906324
  
  
  Stop_2
   Neutralino_1 t-quark         0.22065736    3.00862752
   Neutralino_2 t-quark         0.63812557    8.70073930
   Neutralino_3 t-quark         0.31053610    4.23410966
   Neutralino_4 t-quark         1.44635779   19.72085545
   Chargino_1 b-quark           1.59111038   21.69453369
   Chargino_2 b-quark           1.47358454   20.09208783
   Stop_1  Z                    1.38482788   18.88190500
   Stop_1  h0                   0.26895379    3.66714154
  Total width :                 7.33415341
  
  
  Chargino_1
   Smuon_1 neutrino             0.00003886    0.28387005
   Stau_1 neutrino              0.01301743   95.10376372
   Neutralino_1 W               0.00055115    4.02662386
   neutralino_1 d^* u           0.00000003    0.00020879
   neutralino_1 s^* c           0.00000003    0.00020863
   neutralino_1 e^+    nu       0.00002675    0.19544840
   neutralino_1 mu^+   nu       0.00002675    0.19544086
   neutralino_1 tau^+  nu       0.00002661    0.19442906
  Total width :                 0.01368761
  
  Chargino_2
   Selectron_2 neutrino         0.12483567    5.01198171
   Smuon_2 neutrino             0.12487681    5.01363320
   Stau_1 neutrino              0.00129050    0.05181199
   Stau_2 neutrino              0.13608583    5.46366024
   e-sneutrino e                0.05066962    2.03431606
   mu-sneutrino mu              0.05073214    2.03682600
   tau-sneutrino tau            0.06846536    2.74879077
   Neutralino_1 W               0.17054406    6.84711094
   Neutralino_2 W               0.71196465   28.58440836
   Chargino_1 Z                 0.60191556   24.16608771
   Chargino_1 h0                0.44847615   18.00570494
   neutralino_1 b^* t           0.00028615    0.01148873
   neutralino_2 d^* u           0.00000457    0.00018361
   neutralino_2 s^* c           0.00000457    0.00018367
   neutralino_2 b^* t           0.00002663    0.00106920
   chargino_1 u u^*             0.00001046    0.00041986
   chargino_1 c c^*             0.00001045    0.00041949
   chargino_1 d d^*             0.00000445    0.00017860
   chargino_1 s s^*             0.00000446    0.00017891
   chargino_1 b b^*             0.00052754    0.02117995
  Total width :                 2.49074476
  
  Neutralino_1 : stable
  
  Neutralino_2
   Selectron^-_1 e^+            0.00067750    3.46314577
   Selectron^+_1 e^-            0.00067750    3.46314577
   Smuon^-_1 mu^+               0.00070088    3.58266441
   Smuon^+_1 mu^-               0.00070088    3.58266441
   Stau^-_1 tau^+               0.00836773   42.77287626
   Stau^+_1 tau^-               0.00836773   42.77287626
   Neutralino_1 photon          0.00000004    0.00019852
   neutralino_1 u u^*           0.00000350    0.01786695
   neutralino_1 c c^*           0.00000349    0.01784001
   neutralino_1 d d^*           0.00000463    0.02368626
   neutralino_1 s s^*           0.00000463    0.02368617
   neutralino_1 b b^*           0.00000476    0.02435045
   neutralino_1 nu_e   nu_e^    0.00001156    0.05908396
   neutralino_1 nu_mu  nu_mu    0.00001156    0.05909905
   neutralino_1 nu_tau nu_ta    0.00001244    0.06357803
   neutralino_1 e^-    e^+      0.00000506    0.02584007
   neutralino_1 mu^-   mu^+     0.00000505    0.02582376
   neutralino_1 tau^-  tau^+    0.00000422    0.02157388
  Total width :                 0.01956317
  
  
  Neutralino_3
   Selectron^-_1 e^+            0.00246761    0.12381862
   Selectron^+_1 e^-            0.00246761    0.12381862
   Selectron^-_2 e^+            0.00110911    0.05565227
   Selectron^+_2 e^-            0.00110911    0.05565227
   Smuon^-_1 mu^+               0.00249046    0.12496504
   Smuon^+_1 mu^-               0.00249046    0.12496504
   Smuon^-_2 mu^+               0.00115204    0.05780683
   Smuon^+_2 mu^-               0.00115204    0.05780683
   Stau^-_1 tau^+               0.01021462    0.51254502
   Stau^+_1 tau^-               0.01021462    0.51254502
   Stau^-_2 tau^+               0.01218195    0.61126086
   Stau^+_2 tau^-               0.01218195    0.61126086
   e-sneutrino nu_e^*           0.00612932    0.30755439
   e-sneutrino^* nu_e           0.00612932    0.30755439
   mu-sneutrino nu_mu^*         0.00612947    0.30756211
   mu-sneutrino^* nu_mu         0.00612947    0.30756211
   tau-sneutrino nu_tau^*       0.00617278    0.30973532
   tau-sneutrino^* nu_tau       0.00617278    0.30973532
   Chargino^+_1 W^-             0.59455684   29.83342506
   Chargino^-_1 W^+             0.59455684   29.83342506
   Neutralino_1 Z               0.22190910   11.13486222
   Neutralino_2 Z               0.41946063   21.04752045
   Neutralino_1 h0              0.04208191    2.11156877
   Neutralino_2 h0              0.02422138    1.21537033
   Neutralino_2 photon          0.00002024    0.00101568
   neutralino_1 b b^*           0.00000772    0.00038719
   neutralino_2 b b^*           0.00000522    0.00026192
  Total width :                 1.99292182
  
  
  Neutralino_4
   Selectron^-_1 e^+            0.01008587    0.38204392
   Selectron^+_1 e^-            0.01008587    0.38204392
   Selectron^-_2 e^+            0.02480502    0.93959259
   Selectron^+_2 e^-            0.02480502    0.93959259
   Smuon^-_1 mu^+               0.01007017    0.38144944
   Smuon^+_1 mu^-               0.01007017    0.38144944
   Smuon^-_2 mu^+               0.02486868    0.94200385
   Smuon^+_2 mu^-               0.02486868    0.94200385
   Stau^-_1 tau^+               0.00789319    0.29898723
   Stau^+_1 tau^-               0.00789319    0.29898723
   Stau^-_2 tau^+               0.04123849    1.56207842
   Stau^+_2 tau^-               0.04123849    1.56207842
   e-sneutrino nu_e^*           0.06494620    2.46010577
   e-sneutrino^* nu_e           0.06494620    2.46010577
   mu-sneutrino nu_mu^*         0.06494764    2.46016018
   mu-sneutrino^* nu_mu         0.06494764    2.46016018
   tau-sneutrino nu_tau^*       0.06535235    2.47549054
   tau-sneutrino^* nu_tau       0.06535235    2.47549054
   Chargino^+_1 W^-             0.67613622   25.61145377
   Chargino^-_1 W^+             0.67613622   25.61145377
   Neutralino_1 Z               0.05545958    2.10076061
   Neutralino_2 Z               0.04990974    1.89053781
   Neutralino_1 h0              0.18282528    6.92526289
   Neutralino_2 h0              0.37096704   14.05190972
   Neutralino_2 photon          0.00002906    0.00110069
   chargino^+_1 d u^*           0.00000417    0.00015812
   chargino^-_1 d^* u           0.00000417    0.00015812
   chargino^+_1 s c^*           0.00000418    0.00015820
   chargino^-_1 s^* c           0.00000418    0.00015820
   chargino^+_1 b t^*           0.00001489    0.00056401
   chargino^-_1 b^* t           0.00001489    0.00056401
   neutralino_1 b b^*           0.00000829    0.00031410
   neutralino_2 u u^*           0.00000460    0.00017408
   neutralino_2 c c^*           0.00000459    0.00017392
   neutralino_2 d d^*           0.00000530    0.00020077
   neutralino_2 s s^*           0.00000530    0.00020085
   neutralino_2 b b^*           0.00001293    0.00048988
  Total width :                 2.63997596
  
  
  Gluino
   Sup_1 u^*                    0.22870900    4.94258582
   Sup_1^* u                    0.22870900    4.94258582
   Sup_2 u^*                    0.11126815    2.40459439
   Sup_2^* u                    0.11126815    2.40459439
   S-charm_1 c^*                0.22850951    4.93827473
   S-charm_1^* c                0.22850951    4.93827473
   S-charm_2 c^*                0.11134187    2.40618760
   S-charm_2^* c                0.11134187    2.40618760
   Stop_1 t^*                   0.23793220    5.14190657
   Stop_1^* t                   0.23793220    5.14190657
   Sdown_1 d^*                  0.23074812    4.98665276
   Sdown_1^* d                  0.23074812    4.98665276
   Sdown_2 d^*                  0.08317483    1.79747505
   Sdown_2^* d                  0.08317483    1.79747505
   S-strange_1 s^*              0.23077130    4.98715370
   S-strange_1^* s              0.23077130    4.98715370
   S-strange_2 s^*              0.08317883    1.79756163
   S-strange_2^* s              0.08317883    1.79756163
   Sbottom_1 b^*                0.51851488   11.20552447
   Sbottom_1^* b                0.51851488   11.20552447
   Sbottom_2 b^*                0.24263650    5.24357028
   Sbottom_2^* b                0.24263650    5.24357028
   Stop_1 c^*                   0.00530170    0.11457405
   Stop_1^* c                   0.00530170    0.11457405
   neutralino_1 gluon           0.00002680    0.00057923
   neutralino_2 gluon           0.00022135    0.00478362
   neutralino_3 gluon           0.00033270    0.00718998
   neutralino_4 gluon           0.00038117    0.00823738
   neutralino_1 t t^*           0.00008962    0.00193684
   neutralino_2 t t^*           0.00009374    0.00202577
   chargino^+_1 t^* b           0.00066796    0.01443507
   chargino^-_1 t b^*           0.00066796    0.01443507
   chargino^+_2 t^* b           0.00032981    0.00712747
   chargino^-_2 t b^*           0.00032981    0.00712747
  Total width :                 4.62731469
  
 h0 
   muons                        0.00000101    0.04633093
   taus                         0.00028643   13.08657852
   d-quark                      0.00000000    0.00016098
   s-quark                      0.00000104    0.04730787
   b-quark                      0.00180832   82.62040292
   c-quark                      0.00009191    4.19919143
  Total width :                 0.00218871
  
  
 H0 
   muons                        0.00027932    0.03671913
   taus                         0.07900846   10.38638120
   d-quark                      0.00000098    0.00012915
   s-quark                      0.00028872    0.03795547
   b-quark                      0.50606084   66.52630492
   c-quark                      0.00000440    0.00057844
   t-quark                      0.03284785    4.31814858
   Selectron 1 1                0.00038588    0.05072782
   Smuon 1 1                    0.00039320    0.05168917
   Smuon 1 2                    0.00001885    0.00247807
   Smuon 2 1                    0.00001885    0.00247807
   Stau 1 1                     0.00420030    0.55216824
   Stau 1 2                     0.00406411    0.53426431
   Stau 2 1                     0.00406411    0.53426431
   e-Sneutrino                  0.00071373    0.09382650
   mu-Sneutrino                 0.00071386    0.09384330
   tau-Sneutrino                0.00074895    0.09845649
   neutralino_1 neutralino_1    0.01640703    2.15685379
   neutralino_1 neutralino_2    0.04741210    6.23275196
   neutralino_2 neutralino_2    0.01321269    1.73692880
   chargino^+_1 chargino^-_1    0.03572858    4.69684775
   Z Z                          0.00145341    0.19106428
   W+ W-                        0.00311215    0.40912109
   h0 h0                        0.00955444    1.25601818
  Total width :                 0.76069285
  
  
  
 A0 
   muons                        0.00027991    0.02394510
   taus                         0.07918234    6.77364982
   s-quark                      0.00028933    0.02475108
   b-quark                      0.50718753   43.38733581
   c-quark                      0.00000337    0.00028798
   t-quark                      0.10446570    8.93651435
   Smuon 1 2                    0.00001955    0.00167248
   Smuon 2 1                    0.00001955    0.00167248
   Stau 1 2                     0.00569243    0.48695874
   Stau 2 1                     0.00569243    0.48695874
   neutralino_1 neutralino_1    0.02519311    2.15514326
   neutralino_1 neutralino_2    0.10809913    9.24733561
   neutralino_2 neutralino_2    0.09248237    7.91140044
   chargino^+_1 chargino^-_1    0.23769243   20.33338946
   h0 Z                         0.00267578    0.22889978
  Total width :                 1.16897596
  
  
  
  H^+
   muon neutrino                0.00028573    0.04266509
   tau neutrino                 0.08082739   12.06922654
   d-quark u-quark              0.00000088    0.00013177
   s-quark c-quark              0.00026233    0.03917183
   b-quark t-quark              0.42729169   63.80362064
   Selectron_2 Sneutrino        0.00051866    0.07744696
   Smuon_1 Sneutrino            0.00005543    0.00827732
   Smuon_2 Sneutrino            0.00051472    0.07685861
   Stau_1 Sneutrino             0.01489058    2.22347616
   Stau_2 Sneutrino             0.00000817    0.00121960
   chargino_1 neutralino_1      0.14117172   21.07990203
   chargino_1 neutralino_2      0.00083132    0.12413279
   h0 W                         0.00303956    0.45386947
  Total width :                 0.66969819
  
  
 Total cross sections in fb for:
  E_cms :     500.0000000  GeV
  Degree of polarization: P_e- = 0.000000 P_e+ = 0.000000
  Intial state radiation is included
  
  u-squarks : kinematically not possible
  
  c-squarks : kinematically not possible
  
  t-squarks : kinematically not possible
  
  
  d-squarks : kinematically not possible
  
  s-squarks : kinematically not possible
  
  b-squarks : kinematically not possible
  
  
  selectrons
   1   1     283.6216345 fb
   1   2      78.0199000 fb
   2   2      45.7859068 fb
  
  smuons
   1   1      54.7929652 fb
   1   2       0.0059693 fb
   2   2      19.0224584 fb
  
  staus
   1   1      59.6056340 fb
   1   2       1.2593673 fb
   2   2      15.9954658 fb
  
  
  e-sneutrino
             453.2694763 fb
  
  mu-sneutrino
              13.6239191 fb
  
  tau-sneutrino
              13.9240765 fb
  
  
  neutralinos
   1   1     275.7338081 fb
   1   2      65.7329474 fb
   1   3       7.2644701 fb
   1   4       0.9005054 fb
   2   2      69.0994560 fb
  
  
  charginos
   1   1     161.0778560 fb
  
  
  h^0 Z
              24.7411048 fb
  
  H^0 Z : cross section below   1.0000000E-03 fb
  
  h^0 A^0 : kinematically not possible
  
  H^0 A^0 : kinematically not possible
  
  H^+ H^- : kinematically not possible
  
\end{verbatim}

\end{appendix}

\end{document}